\documentclass[twocolumn,pra,showpacs,preprintnumbers,amsfonts,amsmath,amssymb]{revtex4}

\usepackage{graphicx}
\usepackage{dcolumn}
\usepackage{bm}

\usepackage{amsthm}

\newtheorem{fact}{Fact}
\newtheorem{thm}{Theorem}
\newtheorem{lem}{Lemma}
\newtheorem{cor}{Corollary}

\theoremstyle{definition}
\newtheorem{defn}{Definition}

\theoremstyle{remark}
\newtheorem*{rem}{Remark}

\newcommand{\su}{\mathfrak{su}}
\newcommand{\SU}{\mathrm{SU}}
\newcommand{\SO}{\mathrm{SO}}
\newcommand{\SP}{\mathrm{Sp}}
\newcommand{\Or}{\mathrm{O}}
\newcommand{\GL}{\mathrm{GL}}
\newcommand{\U}{\mathrm{U}}
\newcommand{\Q}{\mathbb{H}}
\newcommand{\spec}{\mathrm{spec}}
\newcommand{\sgn}{\mathrm{sgn}}

\newcommand{\R}{\mathbb{R}}
\newcommand{\Z}{\mathbb{Z}}
\newcommand{\N}{\mathbb{N}}
\newcommand{\C}{\mathbb{C}}
\newcommand{\G}{\mathrm{G}}
\newcommand{\GA}{\mathrm{A}}

\newcommand{\GK}{\mathrm{K}}

\newcommand{\GP}{\mathrm{P}}
\newcommand{\g}{\mathfrak{g}}
\newcommand{\gh}{\mathfrak{h}}
\newcommand{\gk}{\mathfrak{k}}
\newcommand{\gp}{\mathfrak{p}}
\newcommand{\ga}{\mathfrak{a}}

\newcommand{\ii}{\dot\imath}
\newcommand{\ad}{\mathrm{ad}}
\newcommand{\Ad}{\mathrm{Ad}}
\newcommand{\Tr}{\mathrm{Tr}}
\newcommand{\cc}{\mathfrak{c}}
\newcommand{\spanr}{{\mathrm{span}}_{\R}}
\newcommand{\V}{\mathrm{V}_{\C}}
\newcommand{\W}{\mathrm{W}_{\C}}
\newcommand{\UC}{\mathrm{U}_{\C}}
\newcommand{\B}{\mathcal{B}}
\newcommand{\HV}{\mathcal{H}}
\newcommand{\id}{\mathrm{id}}
\newcommand{\diag}{\mathrm{diag}}
\newcommand{\We}{\mathcal{W}}

\begin{document}

\title{Gate simulation and lower bounds on the simulation time}

\author{Robert Zeier}%
\email{zeier@ira.uka.de}
\author{Markus Grassl}%
\email{grassl@ira.uka.de}
\author{Thomas Beth}%
\email{EISS_Office@ira.uka.de}
\affiliation{%
Institut f{\"u}r Algorithmen und Kognitive Systeme, Universit{\"a}t
Karlsruhe, 76128 Karlsruhe, Germany
}%

\date{March 10, 2004}


\begin{abstract}
Unitary operations are the building blocks of quantum programs. 
Our task is to design efficient or optimal
implementations of these unitary operations by employing the
intrinsic physical resources of a given $n$-qubit system. 
The most common versions of this task are known as Hamiltonian simulation
and gate simulation, where Hamiltonian simulation can be seen as an 
infinitesimal version of the general task of gate simulation.
We present a Lie-theoretic approach to Hamiltonian simulation and
gate simulation. From this, we derive lower bounds
on the time complexity in the $n$-qubit case, generalizing known results
to both even and odd $n$. To achieve this we develop a generalization
of the so-called magic basis for two-qubits. As a corollary,
we note a  connection to entanglement measures
of concurrence-type.
\end{abstract}


\maketitle

\section{Introduction \label{introduction}}
As a starting point for the emerging field of quantum 
computation (for a review see, e.g.,
Refs.~\cite{ABH:2001,NC:2000}) Feynman~\cite{Fey:1982}
established in 1982 a connection between simulation of quantum
systems by computers on the one hand and computation using
quantum systems on the other hand. Since this time many different
models for computation using quantum systems have been proposed.

In all this models the common target is to constructively
implement unitary operations by employing the intrinsic physical 
resources. Forming the counterpart of operations on a classical computer,
these unitary operations build the components of quantum programs. 
Whichever (reasonable) system
is provided to us by an experimentalist, we
exploit the resources immanent to the system. 

In this paper we confine ourselves to $n$-qubit systems where the 
resources are given by local unitary operations and by the natural 
time evolution specified by the Hamilton operator. 
This allows us to implement a given unitary operation
by interrupting the natural time evolution
with local unitary operations.
Referring to such implementations as programs,
our objective is to study efficient or optimal programs.
To achieve this, usually two different versions of this problem 
are considered. 
Hamiltonian simulation is the one version and denotes an 
infinitesimal implementation of unitary operations, i.e.,
the unitary operation is in a neighborhood of the identity.
The second version is called gate simulation and describes an 
implementation of unitary operations not restricted 
to a  neighborhood of the identity.
We give an exact definition of  both versions in Sec.~\ref{model}.

In this paper we address two major topics. First
we consider two-qubit systems. Combining the manifold results known
for two-qubit systems, we can close gaps and simplify 
the line of reasoning. In addition, by using Lie-theoretic
methods we derive a unified approach to the methodology. The motivation
for this extensive reconsideration is twofold: First,
we show that understanding simple cases in detail helps 
to generalize them to higher-dimensional systems.
Secondly, we obtain generally applicable (Lie-theoretic) 
methods which provide tools in the analysis of
higher-dimensional systems. 

Beyond this, the two-qubit case is interesting in its own. 
We can characterize the minimal time for Hamiltonian simulation
directly using arguments from Lie theory. With the help
of this characterization we explain and reprove a majorization-like 
condition \cite{BCL:02} for the minimal simulation time. By employing the 
Weyl group of the corresponding Lie algebra, we are able to
simplify and clarify the known approach of Ref.~\cite{BCL:02},
especially w.r.t.\ Ref.~\cite{KBG:2001}.
In the case of two-qubit systems we consider gate simulation as well.
We present a refined analysis of the majorization-like condition 
of Refs.~\cite{VHC:2002,HVC:2002},
which will be built explicitly on the results of Ref.~\cite{KBG:2001}. 

The second major topic deals with the case of general $n$-qubit systems.
We first discuss a  generalization of the so-called magic
basis \cite{BDS:1996,HW:1997} to higher dimensional systems. 
With this information on the structure of 
unitary operations we
develop lower bounds on the minimal time for gate simulations.
Our method applies to all $n$-qubit systems and generalizes
a result of Ref.~\cite{CHN:2003} for even $n$. We also discuss
the used techniques in connection to entanglement measures, in
particular, the concurrence~\cite{HW:1997,Woo:1998}.

The whole text is written in a Lie-theoretic flavor. 
For further reference, the Lie-theoretic 
concepts needed in the text will be introduced briefly in 
Sec.~\ref{lieprep}. 
Using this theory puts us in the position to formulate
strong arguments in a coherent language. 

Firstly, in Sec.~\ref{model} we introduce our model.
In Sec.~\ref{lieprep} we state the Lie-theoretic concepts
needed in the main body of the text. The Hamiltonian simulation
for two-qubits will be discussed in Sec.~\ref{infini}, followed by
the analysis of gate simulation for two-qubits in Sec.~\ref{gatesim}. 
The generalization
of the magic basis for two-qubits is considered together with
lower bounds on the time complexity for general $n$-qubit
systems in Sec.~\ref{sectlow}. 
In Sec.~\ref{relwork} we give a brief outline to related work,
in Sec.~\ref{dis}
we continue with a discussion
of connections of our approach to 
concurrence-type entanglement measures,
and 
in Sec.~\ref{conclusion}
we close with the conclusion. In the appendix we recall
a spectral approach to infinitesimal Hamiltonian simulation.

\section{The model}\label{model}
We consider a system of $n$ qubits, where $n\in {\N}$ is
finite. This system can be modeled within an 
$n$-fold  tensor product ${\C}^2\otimes \cdots \otimes 
{\C}^2$ of $n$ two-dimensional complex vector spaces
${\C}^2$, i.e., a tensor product of single qubits. 
The time evolution of the system is governed by the
Schr{\"o}dinger equation for the time-evolution operator, see, e.g., 
Ref.~\cite[p.~72]{Sak:1994},
\begin{equation*}
\frac{d}{d t} U(t) = (- \ii H) U(t),
\end{equation*}
where $t$ denotes the time, $U(t)$ the time evolution operator, and $H$ the
Hamilton operator which is supposed to be time-independent
($\hbar=1$).  Because of the
irrelevance of a global phase in quantum mechanics, we restrict ourselves 
to evolution operators from the special unitary group $\SU(2^n)$.

In addition to the possibility to let the system evolve according to
the evolution operator $U(t)=\exp(- \ii H t)$, 
in our model we allow  the application of local unitary operators. 
A unitary operator is considered as local when it does not induce any 
interaction between different qubits, i.e., when it has the form of an 
$n$-fold tensor product $U_{1}\otimes \cdots \otimes U_{n}$ of 
unitary operators $U_{i} \in \SU(2)$ with $i \in \{1,\ldots,n\}$. 
The time for the application of local unitary operations
is negligible and assumed to be zero. Thereby we have specified the 
available resources  which 
constitute the  possibilities to control the system. 

We emphasize that, for mathematical reasons, we restrict our model 
to consider only systems 
for which the system Hamilton operators can be represented
without use of local terms. As a consequence, in the case of infinitesimal
Hamiltonian simulation we can simulate Hamilton operators exactly
only if the Hamilton operator can be represented without 
use of local terms. To remove the 
local terms of a system Hamilton operator one usually employs some
approximations, see, e.g., Ref.~\cite[p.~3]{BCL:02} or
Ref.~\cite[p.~288]{MVL:02}. 
But we refrain from considering such approximations.
In addition, avoiding local terms in Hamilton
operators seems to release  us from some problems
with infinite programs, i.e., an infinite
number $m=\infty$ of steps (see below). 
In Ref.~\cite{HNO:2003} it is analyzed under which
conditions we need infinite programs for time optimal control in the
more general setting which includes local terms in the system Hamilton 
operator.  Nevertheless, we consider for technical reasons all types
of programs, even infinite ones.
The available resources will be utilized below in three ways.

First, we consider the simulation of a unitary gate, i.e., a unitary
operator. This
notion of simulation means that the system is able to implement
a given unitary gate by interrupting the natural time evolution
with local unitary operations.
Definition~\ref{defn gate simulation}
states this in a more formal way. 
The term ``gate simulation'' was introduced
in Ref.~\cite[p.~3]{BCL:02}.

\begin{defn}[Gate simulation\label{defn gate simulation}]
An $n$-qubit system with Hamilton operator $H$ and local unitary
operators available simulates  a unitary gate $U$ in time $t$ 
if there exists
local unitary
operators $U_{0}$ and $U_{j}$ as well as
times $t_{j}\geq 0$ with $t=\sum_{j=1}^{m} t_{j}$, $j \in {\N}$, 
$1\leq j \leq m$, and $m \in
{\N}\cup \{0,\infty \}$, so that
\begin{equation*}
U = \left[ \prod_{j=1}^{m} U_{j} \exp(- \ii H t_{j}) \right] U_{0}.
\end{equation*}
\end{defn}

\begin{rem}
Due to the non-commutativity of the unitary group we restrict the
symbol $\prod$ for elements $V_{j}$ from the unitary group  to the 
following meaning:
\begin{equation*}
\prod_{j=e}^{f} V_{j} := 
\left\{
\begin{aligned}
&\left( \prod_{j=e+1}^{f} V_{j} \right) V_{e}& \text{for $f \geq e$},\\
&\id& \text{for $f < e$},  
\end{aligned}
\right.
\end{equation*}
where $e,f \in {\Z}$, and the identity element of the unitary group
is denoted by $\id$. In the case of $f$ infinite 
the symbol~$\prod$ represents an element from the closure of
convergent sequences.
\end{rem}

Secondly, we introduce a particular concept of infinitesimal simulation of
a unitary gate. A given unitary gate $U$ is treated as a point of 
a one-parameter group $\exp(- \ii H' t')$.
The infinitesimal simulation of the Hamilton operator $H'$ denotes
that the system simulates the corresponding one-parameter group 
for infinitesimal  times $t'$, i.e.,
the derivatives of the one-parameter group and of the simulation
coincide for infinitesimal times. 
We emphasize that in Def.~\ref{defn Hsim}
the notion of infinitesimal Hamiltonian simulation  
is defined independently of the unitary gate $U$.

\begin{defn}[Infinitesimal Hamiltonian simulation\label{defn Hsim}]
An $n$-qubit system with Hamilton operator $H$ and  local unitary
operators available simulates
an Hamilton operator $H'$ infinitesimally in time $t$ if  there exists
local unitary operators $U_{0}$ and $U_{j}$   
as well as  
times $t_{j}\geq 0$ where $t=\sum_{j=1}^{m} t_{j}$, $j \in {\N}$, 
$1\leq j \leq m$, and $m \in
{\N}\cup \{0,\infty \}$ 
so that 
$\prod_{j=0}^{m} U_{j}$ is equal to the identity  of 
the unitary group and 
the following equation holds:
\begin{multline}\label{eq ihs}
\lim_{\substack{t' \to 0\\ t'>0}} \left[ \frac{d}{dt'} 
\exp(- \ii t' H')  \right] = \\ 
\lim_{\substack{t' \to 0\\ t'>0}} \left( \frac{d}{dt'} \left\{ \left[
\prod_{j=1}^{m} U_{j} \exp(-
\ii  t'  H t_{j}) \right] U_{0}\right\} \right).
\end{multline}
\end{defn}

\begin{rem}
The condition $\prod_{j=0}^{m} U_{j}= \id$ ensures that our program
specified by $U_{0}$, the $U_{j}$s, and the $t_{j}$s operates 
nearby the identity for $t' \to 0$ .
\end{rem}

We note that in Refs.~\cite{WJB:02,WJB:02b,WRJB:02,WRJB:02b} the
notion of infinitesimal Hamiltonian simulation was extended
by the so-called first order approximation to unitary operators.
Similar ideas were used in Ref.~\cite{BCL:02} where 
it was proposed to follow the evolution of the system exactly by
the Hamiltonian simulation. Concurrently, it was remarked in 
Ref.~\cite{BCL:02} that to follow the evolution of system exactly
is only possible infinitesimally,
as the control is not continuous. In this text we do not consider
such approximations.

Though Def.~\ref{defn Hsim} presents the essential meaning of
infinitesimal Hamiltonian simulation, it  appears to be very unpractical.
Thus we present an equivalent condition, which is usually formulated as 
definition~\cite{WJB:02,WJB:02b,WRJB:02,WRJB:02b,Chen:2003}.

\begin{lem}
An $n$-qubit system with Hamilton operator $H$ and  local unitary
operators available simulates
the Hamilton operator $H'$ infinitesimally in time $t$ if and only if
there exists local unitary operators $V_{j}$ as well as 
times $t_{j}\geq 0$ with $t=\sum_{j=1}^{m} t_{j}$, $j \in {\N}$, 
$1\leq j \leq m$, and $m \in
{\N}\cup \{0,\infty \}$ so that 
the following equation holds:
\begin{equation}\label{fact ihs}
H' =  \sum_{j=1}^{m} t_{j} ( {V_{j}}^{-1} H {V_{j}}).
\end{equation}
\end{lem}

\begin{proof}
The ``only if''-case:
The l.h.s.\ of Eq.~(\ref{eq ihs}) equals $-\ii H'$ and the r.h.s. of
Eq.~(\ref{eq ihs}) equals
\begin{equation}\label{pr ihs}
\lim_{\substack{t' \to 0\\ t'>0}} \left[ \frac{d}{dt'} \left( \left\{
\prod_{j=1}^{m} \exp\left[-
\ii  t' ( {W}_{j} H {{W}_{j}}^{-1} ) t_{j}\right] \right\} {W}_{0}\right) \right],
\end{equation}
where
\begin{equation*}
{W}_{j} = 
\begin{cases}
U_{m}& \text{for $j=m$},\\ 
{W}_{j+1}  U_{j}& \text{for $0\leq j <m$}.
\end{cases}
\end{equation*}
We differentiate, compute the limit, and equate 
the result of Eq.~(\ref{pr ihs}) with $-\ii H'$. After this we
use $V_{j}:={W_{j}}^{-1}$ and
$W_{0}=\prod_{j=0}^{m} U_{j}=\id$ to obtain Eq.~(\ref{fact ihs}).

The ``if''-case: After insertion of Eq.~(\ref{fact ihs}) in
Eq.~(\ref{eq ihs}) we obtain the ``if''-case.
\end{proof}

Thirdly, we introduce in Def.~\ref{defn infinitesimal gate sim} 
the notion of infinitesimal gate simulation
which depends explicitly on the given unitary gate $U$.
\begin{defn}[Infinitesimal gate 
simulation\label{defn infinitesimal gate sim}]
An $n$-qubit system with Hamilton operator $H$ and local unitary
operators available simulates  a unitary gate $U$ infinitesimally 
in time $t$ if there exists an Hamilton operator $H'$ and local 
unitary gates $U_{1}$ and $U_{2}$ 
such that the  equation $U = U_{1} \exp(- \ii H') U_{2}$ holds
and the system simulates the Hamilton operator $H'$
infinitesimally in time $t$.
\end{defn}
\begin{rem}
Actually, we do not use Def.~\ref{defn infinitesimal gate sim}
later in the text. But we state this definition to highlight
that Def.~\ref{defn Hsim} is independent of some unitary
operator $U$ and does not incorporate different
decompositions of $U$ which could lead to different Hamilton operators
$H'$. The existence of different Hamilton operators $H'$
will be employed in Sec.~\ref{gatesim} below.
\end{rem}

Before we proceed, we discuss our model.
Entanglement describes important non-local properties of states and gates.
Because entanglement is invariant under local unitary
operations~\cite{NC:2000} it seems reasonable to neglect the time
needed to implement local unitary gates for the
implementation of general unitary gates. This is supported by the fact 
that two-qubit gates are considered as significantly more difficult
to implement than one-qubit gates~\cite{ROADMAP:2002}.
Additionally, in Nuclear Magnetic Resonance (NMR) the application 
of local unitary operations in zero time is expressed
by the notion of the ``fast control limit'' which
is conventionally considered as a good approximation. This is 
reasonable because local and non-local gates operate on
different time scales~\cite{HW:1968,EBW:1997,KBG:2001}.

\section{Lie-theoretic preparations\label{lieprep}}
In this section we recall Lie-theoretic notions and methods
which will be employed throughout the paper.  This reflects the
intimate connection of the considered problems to Lie theory
and it makes the text more readable and self-contained.
For convenience, this section can also be regarded as a reference section. 
We remark that our 
presentation in this section was partly inspired by 
Refs.~\cite{KBG:2001,KG:2001b,ZVSW:2003}.

\subsection{Basic concepts}
For our purposes we can consider Lie groups as linear matrix groups, i.e.,
as closed subgroups of the general linear group. The tangent space
to the Lie group $\G$ at the identity is isomorphic to the Lie algebra
$\g$ corresponding to $\G$. A Lie algebra, 
which is in particular a vector space, comes with a 
bilinear and skew-symmetric multiplication
operation called the Lie bracket $[ \; , \; ]$. The Lie algebra
$\g$ is closed under the Lie bracket and
the Jacobi identity
\begin{equation*}
[[g_{1},g_{2}],g_{3}]
+
[[g_{3},g_{1}],g_{2}]
+
[[g_{2},g_{3}],g_{1}]=0
\end{equation*}
holds for all elements $g_{1}, g_{2}, g_{3} \in \g$. 
We emphasize that we use only  real or complex Lie algebras
which are finite-dimensional.
For general reference on
Lie groups and Lie algebras
please consult Refs.~\cite{Hel:2001,Sam:1990,
Kna:2002,Var:1984,KN:1991,KN:1996,Wol:1984,Loo:1969a,Loo:1969b,Bor:1998,Gil:1994}.
In the following
let $\G$ denote a Lie group and $\g$ its associated
Lie algebra. 

The map $\ad_{\g}(g)$ from the Lie algebra $\g$ to itself
is defined by $h \mapsto [g,h]$, where  
$h,g \in \g$. With this notation, the adjoint 
representation of the Lie algebra $\g$ in itself 
is given by $g \mapsto \ad_{\g}(g)$. Let $\ad_{\g}(\gh)$
denote the set $\{\ad_{\g}(h)|\, h \in \gh \}$ for some subspace
$\gh$ of the Lie algebra $\g$.
Now, we can introduce a symmetric
bilinear form on the Lie algebra $\g$: the Killing form
$B_{\g}(g,h):= \Tr_{\g} (\ad_{\g}(g) \circ \ad_{\g}(h))$. 
If the Killing form is non-degenerate it
can be thought as an inner product on the Lie algebra,
although it does in general  not fulfill the axiom
of positivity.

\begin{defn}[Orthogonal symmetric Lie algebra, see 
Ref.~{\cite[p.~213]{Hel:2001}} and 
Ref.~{\cite[pp.~225--226 and 246]{KN:1996}}\label{osla}] 
A pair $(\g,\theta)$ is an orthogonal symmetric Lie algebra if
\begin{enumerate}
\item[(i)] $\g$ is a real Lie algebra,
\item[(ii)] $\theta$ is an involutive automorphism of $\g$,
\item[(iii)] and the connected Lie group of linear transformations
of $\g$ generated by $\ad_{\g}(\gk)$ is compact, where $\gk$ is
the set of fixed points of $\theta$ in $\g$.
\end{enumerate}
\end{defn}

\begin{rem}
An automorphism of a Lie algebra $\g$ respects the Lie bracket, i.e.,
for all $g$ and $h$ in $\g$ we have
$\theta([g,h])=[\theta(g),\theta(h)]$. An involutive automorphism is
in addition self-inverse.
 Let $\gk$ and $\gp$ be the eigenspaces of $\theta$ in $\g$ for
the $+1$ and $-1$ eigenvalue respectively. 
Consider the canonical decomposition $\g = \gk + \gp$.
Condition (ii) of
Def.~\ref{osla} is equivalent to
\begin{equation}\label{komm}
[ \gk , \gk ] \subset \gk,\; [ \gk , \gp ] \subset \gp,
\; [ \gp , \gp ] \subset \gk 
\text{ (see \cite[p.~226--227]{KN:1996}).}
\end{equation}
In Ref.~\cite{KBG:2001} this decomposition was called 
Cartan decomposition.
If Eq.~(\ref{komm}) holds we can define 
$\theta$ by
\begin{equation}\label{involutive}
\theta(k)=k \text{ for all } k \in \gk \text{ and } \theta(p)=-p 
\text{ for all } p \in \gp. 
\end{equation}
In addition when $\g$ is the Lie algebra of a compact group  $\G$ then
Condition (iii) of Def.~\ref{osla} is always true.
\end{rem}

For further reference, we assume that $\g$ is semisimple, i.e.,
that the Killing form of $\g$ is non-degenerate,
and that
$(\g,\theta)$ is an orthogonal symmetric Lie algebra. 
As an example, the Lie algebra $\su(2^n)$, which corresponds
to the Lie group $\SU(2^n)$, is semisimple. We fix  a
canonical decomposition $\g = \gk + \gp$
satisfying Eq.~(\ref{komm}) and a maximal Abelian subalgebra $\ga$
contained in $\gp$. Let $\GK=\exp(\gk)$ and $\GA=\exp(\ga)$ 
denote the subgroups of $\G$ generated by $\gk$ and $\ga$ respectively.
After this preparation we obtain a decomposition $\G = \GK\, \GA\, \GK$
of the Lie group $\G$.

\begin{fact}[$\GK\, \GA\, \GK$ decomposition of the Lie group
$\G$ {\cite[Ch. V, Thm. 6.7]{Hel:2001}}\label{KAK}]
With the notation as given above, the Lie group $\G$ corresponding
to $\g$ can be decomposed as
\begin{equation*}
\G = \GK\, \GA\, \GK.
\end{equation*}
\end{fact}

Similar to the adjoint representation $\ad_{\g}$ of a Lie algebra $\g$ 
in itself,
we can define the adjoint representation $\Ad_{\g}$ of a Lie group $\G$
in its Lie algebra $\g$. For an element $G \in  \G$ we  introduce
 $\phi_{\G}(G)$  as the map $H \mapsto G^{-1}
H G$ with the signature $\G \to \G$. The map $\Ad_{\g}(G)$ has the signature
$\g \to \g$ and is defined as the differential of $\phi_{\G}(G)$.
For matrix representations we can write $\Ad_{\g}(G)$ as the
map $g \mapsto G^{-1} g G$.
We use the shortcut $\Ad_{\g}(\GK) := \bigcup_{K \in \GK} 
\Ad_{\g}(K)$ and get the relation between 
the subspace $\gp$ and
its Abelian subalgebra
$\ga$:

\begin{fact}[{\cite[Ch. V, Lemma 6.3 (iii)]{Hel:2001}}\label{p char}]
The following equation holds:
\begin{equation*}
\gp = (\Ad_{\g}(\GK))(\ga) .
\end{equation*}
\end{fact}

\subsection{The Weyl group and infinitesimal convexity}
We use the notation
\begin{equation*}
C_{\GK}(\ga):=\{K \in \GK |\, (\Ad_{\g}(K))(a) = a \text{ for all } 
a \in \ga \}
\end{equation*}
and
\begin{equation*}
N_{\GK}(\ga):=\{K \in \GK |\, (\Ad_{\g}(K))(\ga) \subset \ga\}
\end{equation*}
respectively for the centralizer $C_{\GK}(\ga)$ and 
the normalizer $N_{\GK}(\ga)$ of $\ga$ in $\GK$.

\begin{defn}[Weyl group, {see \cite[p.~284]{Hel:2001} or \cite[p.~381]{Kna:2002}}\label{defnweyl}]
The Weyl group corresponding to $\ga$ is the factor group
$N_{\GK}(\ga) / C_{\GK}(\ga)$. We denote this group
by $\We(\G,\GA)$, where $\GA=\exp(\ga)$.
\end{defn}

The Weyl group $\We(\G,\GA)$ is finite (see Fact~\ref{weylrefl} below). 
In order to compute the Weyl group, we  
introduce the concept of restricted roots.

\begin{defn}[Restricted root, cf. {\cite[p.~370]{Kna:2002}}\label{restroots}]
Let $\lambda$ be a linear function on $\ga$.
The linear subspace $\g^{\lambda}$ is given by
\begin{equation*}
\g^{\lambda} = \{ g \in \g |\, [a,g] = \lambda(a) g \text{ for all } a
\in \ga\}.
\end{equation*} 
The linear function $\lambda$ 
is called a restricted root of $\g$ w.r.t.\ $\ga$ if $\g^{\lambda} \neq
\{0\}$ and $\lambda$ is not identically zero on $\ga$.
Let $\Delta_{\ga}$  denote the set of restricted roots of $\g$ w.r.t.\ $\ga$.
\end{defn}

\begin{rem}
In Ref.~\cite[p.~370]{Kna:2002} the restricted roots are defined
w.r.t.\ $\ii \ga$. But the concept of restricted roots
can also be defined w.r.t.\ $\ga$.
\end{rem}

Due to the fact that $\g$ is semisimple
we deduce that the Killing form $B_{\g}$ restricted to $\ga \times \ga$ 
is non-degenerate.
With this in mind, a restricted root $\lambda$ 
is equal to the map
$a \mapsto B_{\g}(a_{\lambda},a)$, where $a_{\lambda} \in \ga$ is uniquely
determined. We extend the Killing form to restricted 
roots by $B_{\g}(\lambda,\mu):=
B_{\g}(a_{\lambda},a_{\mu})$. For every  $\lambda \in \Delta_{\ga}$
the reflection $s_{\lambda}(\mu)$
of a restricted root $\mu \in  \Delta_{\ga}$
w.r.t.\ the hyperplane
$\{a \in \ga |\, \lambda(a)=0\}$ is given by
\begin{equation*}
s_{\lambda}(\mu):=\mu-2 \frac{B_{\g}(\mu,\lambda)}{B_{\g}(\lambda,\lambda)}
\lambda.
\end{equation*}
Following Ref.~\cite[p.~286]{Hel:2001} the reflection
$s_{\lambda}$ can be extended to elements of $\ga$. For
$a \in \ga$ the reflection $s_{\lambda}(a)$
of $a \in \ga$ in the 
hyperplane $\{a \in \ga |\, \lambda(a)=0\}$ is given
by
\begin{equation}\label{weylrefl2}
s_{\lambda}(a)=a-2\frac{B_{\g}(a,a_{\lambda})}{B_{\g}(a_{\lambda},a_{\lambda})}a_{\lambda}.
\end{equation}

With this preparation we get a possibility
to compute the Weyl group
corresponding to $\ga$:
\begin{fact}[{\cite[p.~383]{Kna:2002}}\label{weylrefl}]
The Weyl group corresponding to $\ga$ is finite and is generated
by  the reflections $s_{\lambda}$, where $\lambda \in  \Delta_{\ga}$.
\end{fact}

Recall that $\We(\G,\GA)$ is a subset of $\GK$ and
operates on $\ga$ by $\Ad_{\g}(K)$, where $K \in \GK$. 
\begin{defn}[Weyl orbit, see, e.g., {\cite[p.~422]{Kos:1973}}\label{weylorbit}]
The Weyl orbit $\We(a)$
of $a \in \ga$ is defined as the set 
$\{ (\Ad_{\g}(W))(a) |\, W \in \We(\G,\GA)\}$. 
By appealing to Fact~\ref{p char}
the Weyl orbit $\We(p)$, for $p \in \gp$, is defined  
as $\We(p):=\We(a)$, where $a \in (\Ad_{\g}(\GK))(p)\cap \ga$.
\end{defn}

To understand that the definition of $\We(p)$ for $p \in \gp$ 
is independent of $a$ and thus well defined we now characterize the 
Weyl orbits in more detail.

\begin{fact}[{\cite[Lemma~7.38]{Kna:2002}}\label{weylorbitlemma}]
Let $(\Ad_{\g}(K))(a)=a'$, where $a,a' \in \ga$ and $K \in \GK$.
Then there exists an element $K' \in N_{\GK}(\ga)$ such that
$(\Ad_{\g}(K'))(a)=a'$.
\end{fact} 

By Fact~\ref{weylorbitlemma} two elements $a$ and $a'$ from the
Weyl orbit $\We(p)$ of $p \in \gp$
are conjugated by an element of the Weyl group,
which proves that the definition 
(Def.~\ref{weylorbit}) of $\We(p)$ is independent of $a$.
This shows in addition that the Weyl orbit $\We(p)$ is equal to 
$\ga \cap (\Ad_{\g}(\GK))(p)$.
Let us denote the convex hull
of the Weyl orbit $\We(p)$ by $\cc(p)$. We state now the infinitesimal version of
Kostant's convexity theorem.

\begin{fact}[Kostant's convexity theorem (infinitesimal version), see 
{\cite[Thm.~8.2]{Kos:1973}} or 
{\cite[Thm.~1]{Hec:1980}}\label{Kostant}]
Let $\Gamma$ be the orthogonal projection of $\gp$ on $\ga$ w.r.t.\ to
the Killing form. For every $p \in \gp$ one obtains 
\begin{equation*}
\Gamma\Bigl((\Ad_{\g}(\GK))(p)\Bigr) = \cc(p).
\end{equation*}
\end{fact}

\begin{rem}
The essential meaning of the infinitesimal version of Kostant's convexity theorem
(Fact~\ref{Kostant}) is that
the projection of $(\Ad_{\g}(\GK))(p)$ to $\ga$ w.r.t.\ the Killing form 
is a convex set
and its extreme points are given by the Weyl orbit $\We(p)$.
\end{rem}

In order to characterize the Weyl orbits in more detail we
introduce additional concepts. The subspace $\ga$ can be
divided into connected components called Weyl chambers.
\begin{defn}[Weyl chamber {\cite[p.~287]{Hel:2001}}]
Let $\lambda$ be any restricted
root of $\g$ w.r.t.\ $\ga$. The hyperplanes 
$\{a \in \ga |\, \lambda(a)=0\}$  divide $\ga$ into finitely many 
connected components excluding their boundary hyperplanes. 
Such a connected component is called Weyl chamber.
The closure of a Weyl chamber including their boundary
hyperplanes is called closed Weyl chamber.
\end{defn}
The Weyl group acts on the Weyl chambers:
\begin{fact}[{\cite[Thm.~2.12, Ch.~VII]{Hel:2001}}\label{permWeyl}]
The Weyl group permutes the Weyl chambers.
\end{fact}

We choose some arbitrary, but fixed, order
on the restricted roots of $\g$ w.r.t.\ $\ga$. Therefore, the
restricted roots can be divided into positive and negative (restricted) 
roots, where positive and negative is defined regarding to the chosen 
order. A restricted root is called fundamental if it is positive and
not a sum of two positive (restricted) roots \cite[p.~59]{Sam:1990}.
Let $\{\alpha_{k}\}\subset \Delta_{\ga}$ be the set of fundamental 
(restricted) roots.
Since the Killing form restricted to $\ga \times \ga$ is
non-degenerate we can define as before for every (restricted) root
$\lambda$ the element $a_{\lambda}\in \ga$ so that 
$B_{\g}(a_{\lambda},a)=\lambda(a)$ for all $a \in \ga$.
The set $\{a \in \ga|\, B_{\g}(a_{\alpha_{k}},a)>0 \text{ for all } \alpha_{k}\}$ is
a Weyl chamber and it is called the fundamental Weyl chamber \cite[p.~61]{Sam:1990}.

\begin{fact}[adapted from {\cite[Prop.~I,
Sec.~2.11]{Sam:1990}}\label{factweylchar}]
Let $\lambda$ and $\mu$ be restricted roots corresponding to
elements $a_{\lambda}$ and $a_{\mu}$ 
of the closed fundamental Weyl chamber, respectively.
The element $a_{\mu}$ lies in the convex hull of the Weyl orbit 
$\We(a_{\lambda})$ of $a_{\lambda}$ if and only if
$\lambda(a)\geq \mu(a)$ for all elements
$a$ of the fundamental Weyl chamber.  The condition
$\lambda(a)\geq \mu(a)$ is equivalent to $B_{\g}(a_{\lambda},a)\geq 
B_{\g}(a_{\mu},a)$.
\end{fact}

\subsection{The two-qubit case\label{su4}}
We now treat the case $\G = \SU(4)$. 
To be more concrete we introduce a
matrix representation for the real semisimple 
Lie algebra $\su(4)$ which corresponds to $\G$.
Let 
\begin{equation*}
\sigma_{x}:=
\begin{pmatrix}
0 & 1 \\
1 & 0
\end{pmatrix},\;
\sigma_{y}:=
\begin{pmatrix}
0 & -\ii \\
\ii & 0
\end{pmatrix},\; \text{and}\;
\sigma_{z}:=
\begin{pmatrix}
1 & 0 \\
0 & -1
\end{pmatrix}
\end{equation*} 
be the the Pauli matrices and set 
\begin{equation*}
\sigma_{0}:=
\begin{pmatrix}
1 & 0 \\
0 & 1
\end{pmatrix}
\end{equation*}
to be the identity matrix. We
identify $\sigma_{x}=\sigma_{1}$,
$\sigma_{y}=\sigma_{2}$, and $\sigma_{z}=\sigma_{3}$
and use the following definitions
\begin{gather*}\label{Xi}
X_{1}:=\frac{i}{2}  \sigma_{0} \otimes \sigma_{1},
X_{2}:=\frac{i}{2}  \sigma_{0} \otimes \sigma_{2},
X_{3}:=\frac{i}{2}  \sigma_{0} \otimes \sigma_{3},\\
X_{4}:=\frac{i}{2}  \sigma_{1} \otimes \sigma_{0},
X_{5}:=\frac{i}{2}  \sigma_{2} \otimes \sigma_{0},
X_{6}:=\frac{i}{2}  \sigma_{3} \otimes \sigma_{0},\\
X_{7}:=\frac{i}{2}  \sigma_{1} \otimes \sigma_{1},
X_{8}:=\frac{i}{2}  \sigma_{2} \otimes \sigma_{2},
X_{9}:=\frac{i}{2}  \sigma_{3} \otimes \sigma_{3},\\
X_{10}:=\frac{i}{2} \sigma_{1} \otimes \sigma_{2},
X_{11}:=\frac{i}{2} \sigma_{1} \otimes \sigma_{3},
X_{12}:=\frac{i}{2} \sigma_{2} \otimes \sigma_{1},\\
X_{13}:=\frac{i}{2} \sigma_{2} \otimes \sigma_{3},
X_{14}:=\frac{i}{2} \sigma_{3} \otimes \sigma_{1},
X_{15}:=\frac{i}{2} \sigma_{3} \otimes \sigma_{2}.
\end{gather*}
The 
standard (or defining) representation of $\su(4)$ is
\begin{equation*}
\g := \su(4) = \spanr \{X_{1},\ldots,X_{15}\},
\end{equation*}
where $\spanr$ denotes the real span.
Let 
\begin{gather*}
\gk:= \spanr \{X_{1}, \ldots, X_{6}\},\\
\gp:= \spanr \{X_{7}, \ldots, X_{15}\},\\
\text{and } \ga:= \spanr \{X_{7}, \ldots, X_{9}\}.
\end{gather*}
With this notation one can easily check that $\gk$ and $\gp$ fulfill
the commutator relations in Eq.~(\ref{komm}). Since the group $\SU(4)$
is compact, the pair $(\g, \theta)$ defines an orthogonal symmetric Lie
algebra, where $\theta$ is given by Eq.~(\ref{involutive}). 
The subspace $\ga$ forms a maximal Abelian subalgebra in $\gp$.
The set of restricted
roots w.r.t.\ $\ga$ can be computed as the eigenvalues of
$\ad_{\g}(c_{1} X_{7}+c_{2} X_{8}+c_{3} X_{9})$:
\begin{multline}\label{roots}
\{ 
\pm \ii (c_{2}-c_{3}), \pm \ii (c_{2}+c_{3}), \pm \ii (c_{1}-c_{3}),\\
\pm \ii (c_{1}+c_{3}), \pm \ii (c_{1}+c_{2}), \pm \ii (c_{1}-c_{2})  
\}.
\end{multline}
We use Eq.~(\ref{weylrefl2}) to obtain a generating set for the Weyl group
(corresponding to $\ga$) as a set of matrices 
\begin{multline}\label{weylmatrices}
\left\{
\begin{pmatrix}
1 & 0 & 0\\
0 & 0 & 1\\
0 & 1 & 0
\end{pmatrix},
\begin{pmatrix}
1 & 0 & 0\\
0 & 0 & -1\\
0 & -1 & 0
\end{pmatrix},
\begin{pmatrix}
0 & 0 & 1\\
0 & 1 & 0\\
1 & 0 & 0
\end{pmatrix},
\right.\\
\left.
\begin{pmatrix}
0 & 0 & -1\\
0 & 1 & 0\\
-1 & 0 & 0
\end{pmatrix},
\begin{pmatrix}
0 & -1 & 0\\
-1 & 0 & 0\\
0 & 0 & 1
\end{pmatrix},
\begin{pmatrix}
0 & 1 & 0\\
1 & 0 & 0\\
0 & 0 & 1
\end{pmatrix}
\right\} ,
\end{multline}
which operate on the vectors
\begin{equation}\label{avector} 
\begin{pmatrix}
1 \\ 0 \\ 0
\end{pmatrix}
\triangleq
X_{7},\;
\begin{pmatrix}
0 \\ 1 \\ 0
\end{pmatrix}
\triangleq
X_{8},\;
\text{and }
\begin{pmatrix}
0 \\ 0 \\ 1
\end{pmatrix}
\triangleq
X_{9}.
\end{equation}

With the notation of Eq.~(\ref{avector}) 
the Killing form restricted to $\ga \times \ga$ is given by
\begin{equation*}
B_{\g}(a,b){\bigr\rvert}_{\ga \times \ga}:=
a^{T}
\begin{pmatrix}
-8& 0& 0\\
0& -8& 0\\
0& 0& -8
\end{pmatrix}
b.
\end{equation*}
Now we give the elements $a_{\lambda} \in \ga$ corresponding to
the restricted roots $\lambda$ in Eq.~(\ref{roots}), 
i.e. elements $a_{\lambda} \in \ga$ such that
$B_{\g}(a_{\lambda},a)=\lambda(a)$ for all $a \in \ga$:
\begin{multline*}
\left\{ 
\frac{\pm \ii}{-8} 
\begin{pmatrix}
0 \\
1 \\
-1 
\end{pmatrix},
\frac{\pm \ii}{-8} 
\begin{pmatrix}
0 \\
1 \\
1 
\end{pmatrix},
\frac{\pm \ii}{-8} 
\begin{pmatrix}
1 \\
0 \\
-1 
\end{pmatrix},
\right.\\
\left.
\frac{\pm \ii}{-8} 
\begin{pmatrix}
1 \\
0 \\
1 
\end{pmatrix},
\frac{\pm \ii}{-8} 
\begin{pmatrix}
1 \\
1 \\
0 
\end{pmatrix},
\frac{\pm \ii}{-8} 
\begin{pmatrix}
1 \\
-1 \\
0 
\end{pmatrix}
\right\}.
\end{multline*}
We have used the basis of Eq.~(\ref{avector}) to represent the
elements $a_{\lambda}$.

In order to present our results in the context of Ref.~\cite{BCL:02}
we choose an order on the (restricted) roots, such that   
the roots of Eq.~(\ref{roots}) which have a plus sign constitute
the positive ones. 
With this convention  for an element
$
d=d_{1}X_{7}
+d_{2}X_{8}
+d_{3}X_{9}
$
of the fundamental Weyl chamber we get
the set of equations
\begin{multline}\label{fundWeyl}
\left\{
d_{2}-d_{3}>0, d_{2}+d_{3}>0, d_{1}-d_{3}>0, 
\right.\\
\left.
d_{1}+d_{3}>0, d_{1}+d_{2}>0, d_{1}-d_{2}>0
\right\},
\end{multline}
where we have identified $(\R,>)$ with $(\ii \R,>)$ by defining
$\ii r_{1} > \ii r_{2} \Leftrightarrow r_{1} >  r_{2}$
for all $r_{1},r_{2} \in \R$.

\section{Infinitesimal Hamiltonian simulation for two qubits}\label{infini}
\subsection{Lie-theoretic explanation\label{subHsim1}}
Following Def.~\ref{defn Hsim} we consider now
infinitesimal Hamiltonian 
simulation for two qubits. We emphasize that
in the two-qubit case
local unitary operations correspond 
to elements of $\GK=\exp(\gk)$. We use the notation
of Sec.~\ref{lieprep}, especially that of Sec.~\ref{su4}.
Since we restrict ourselves to Hamilton operators
without local terms (see Sec.~\ref{model}),
we have for all non-local Hamilton operators $H$ and $H'$
that $\ii H \in \gp$ and $\ii H' \in \gp$, where $\gp$ 
is the subvector space of the Lie algebra $\g$ introduced
in Sec.~\ref{lieprep}.
Thus, we can use Fact~\ref{p char} to write every non-local 
Hamilton operator $H'$ as $H'=(\Ad_{\g}((L')^{-1}))(a')$, where
$a'$ is an element of $\ga$ and $L'$ is a local unitary operator.

\begin{thm}\label{thm iHsim2}
Assume that $H$ and $H'$ are non-local Hamilton operators
acting on a two-qubit system.
Let $a'$ be an element  of $\ga$, where
$a'=(\Ad_{\g}(L'))(H')$ for some local unitary operator $L'$.

A two-qubit system with Hamilton operator $H$ and local unitary
operators available is able to simulate the Hamilton operator $H'$ 
in time $t$ if and only if the Hamilton operator $(a'/t)$ 
lies in the convex closure
of the Weyl orbit $\We(H)$ of $H$. The condition is independent
of the choice of $a'$.
\end{thm}

\begin{rem}
Actually, Thm.~\ref{thm iHsim2} is the infinitesimal version of
Fact~\ref{KBGthm} (see below and Ref.~\cite{KBG:2001}). 
In order to clarify
the connection of Thm.~\ref{factHsimmaj} to the work of Ref.~\cite{KBG:2001},
we give here a proof of this infinitesimal version using
arguments of Refs.~\cite{KBG:2001,BCL:02}.
\end{rem}
\begin{proof}
Assume that
\begin{equation}\label{beweq2} 
t \sum_{i=1}^{m_{3}} q''_{i} (\Ad_{\g}(K''_{i}))(H) 
\end{equation}
is a 
simulation of $H'$ in time $t$, where 
$K''_{i}$ are local unitary operators, $H, H' \in \gp$, $q''_{i}\geq 0$, and 
$\sum_{i=1}^{m_{3}} q''_{i} = 1$.
 
Due to Fact~\ref{p char}
there exists $a$ and $a'$ in $\ga$, where
$a=(\Ad_{\g}(L))(H)$ and $a'=(\Ad_{\g}(L'))(H')$
for some 
local unitary operators $L$ and $L'$.
We remark that local unitary operations cost no time.
Thus, the existence of the simulation
in Eq.~(\ref{beweq2}) is equivalent
to the existence of a simulation 
$t \sum_{i=1}^{m_{2}} q'_{i} (\Ad_{\g}(K'_{i}))(a)$ 
of $a'$ by $a$ in time $t$, where 
$K'_{i}$ are some local unitary operators,
$q'_{i}\geq 0$,
and 
$\sum_{i=1}^{m_{2}} q'_{i} = 1$.
Let $\Gamma$ and $\Gamma'$ 
denote the orthogonal projections (w.r.t.\ 
the Killing form) of $\gp$ on $\ga$ and ${\ga}^{\perp}$ respectively. 
We can write the simulation as
\begin{equation*}
t \sum_{i=1}^{m_{2}} q'_{i} \left[\Gamma\Bigl((\Ad_{\g}(K'_{i}))(a)\Bigr) + 
\Gamma'\Bigl((\Ad_{\g}(K'_{i}))(a)\Bigr)\right] = a'.
\end{equation*}
This is equivalent to
\begin{equation}\label{beweq1}
t \sum_{i=1}^{m} q_{i} \Gamma\Bigl((\Ad_{\g}(K_{i}))(a)\Bigr) = a',
\end{equation}
where 
$K_{i}$ are some local unitary operators,
$q_{i}\geq 0$,
and 
$\sum_{i=1}^{m} q_{i} = 1$.
The last equivalence follows in the backward direction
by employing Fact~\ref{Kostant} to rewrite the projection $\Gamma$
as a convex combination and in the forward direction by
the fact that the term 
$
\sum_{i=1}^{m_{2}} q'_{i} \Gamma'\bigl((\Ad_{\g}(K'_{i}))(a)\bigr)
$
has to
be zero.

By the remark following Fact~\ref{Kostant} we 
know that the projection of $(\Ad_{\g}(\GK))(p)$ 
to $\ga$ w.r.t.\ the Killing form 
is a convex set. Thus, we can write Eq.~(\ref{beweq1}) as 
\begin{equation*}
\Gamma\Bigl((\Ad_{\g}(K'))(a)\Bigr) = (a'/t)
\end{equation*}
for some local unitary
operator $K'$.
With Fact~\ref{Kostant} we get that $(a'/t)$ lies 
in the convex closure of the Weyl orbit
$\We(a)$ of $a$. 
Since $a=(\Ad_{\g}(L))(H)$,
we can replace $a$ by $H$ in the preceding sentence. This
proves the theorem except for the independence of the 
choice of $a'$.

Assume that we replace $a'$ by $a'' \in \ga$, 
where $a''=(\Ad_{\g}(L''))(H')$ for some
local unitary operator $L''$. Due to Fact~\ref{weylorbitlemma},
there exist an 
element $W \in \We(\G,\GA)$ so that $a''=(\Ad_{\g}(W^{-1}))(a')$.
Since operating with an element of the Weyl group leaves 
the Weyl orbit $\We(H)$ unchanged, the Weyl orbit  is equal 
to $(\Ad_{\g}(W))(\We(H))$. It is obvious
that the convex closure of the Weyl orbit $\We(H)$ is left
unchanged as well.
Hence, the element $a'$ is in the convex closure of the 
Weyl orbit $\We(H)$ if and only if
$a''$ is.
\end{proof}

For $a' \in \ga$ it was also proven in Ref.~\cite{BCL:02} 
that the set of Hamilton operators $(a'/t)$ which can be simulated 
in time one is convex. We emphasize that the extreme points of this set 
are given by the Weyl orbit $\We(H)$, which can be computed by means of 
Eq.~(\ref{weylmatrices}). In Ref.~\cite{BCL:02} the extreme points were 
given and their extremality was proven 
by  another method. As in Ref.~\cite{BCL:02}, we state now a version
of Thm.~\ref{thm iHsim2} which gives a condition 
for infinitesimal Hamiltonian simulation in the two-qubit case
that is easier to check.
\begin{thm}[{\cite[p.~11]{BCL:02}}\label{factHsimmaj}]
Assume that $H$ and $H'$ are non-local Hamilton operators
acting on a two-qubit system.
Let $a$ and $a'$ be  elements of the closed fundamental Weyl chamber,
where
$a=a_{1}X_{7}+a_{2}X_{8}+a_{3}X_{9}=(\Ad_{\g}(L))(H)$
and 
$a'=a'_{1}X_{7}+a'_{2}X_{8}+a'_{3}X_{9}=(\Ad_{\g}(L'))(H')$
for some local unitary operators $L$ and $L'$.

A two-qubit system with Hamilton operator $H$ and local unitary
operators available is able to simulate the Hamilton operator $H'$ 
in time $t$ iff the following equations hold:
\begin{subequations}\label{smajbed}
\begin{align}
a_{1} & \geq  a'_{1}/t,\\
a_{1} + a_{2} + a_{3} & \geq (a'_{1} + a'_{2} + a'_{3})/t,\\
a_{1} + a_{2} - a_{3} & \geq  (a'_{1} + a'_{2} - a'_{3})/t.
\end{align}
\end{subequations}
\end{thm}

\begin{rem}
We force $a$ and $a'$ to be (almost) unique elements of $\ga$
by choosing them to be elements
of the closed fundamental Weyl chamber. If $a$ or $a'$ lies on the boundary
of the fundamental Weyl chamber, they are elements of
the closed fundamental Weyl chamber, but not elements of 
the fundamental Weyl chamber. Only in this case there remains
a non-uniqueness, and the considered element
can possibly chosen to lie on different boundary hyperplanes
of the closed fundamental Weyl chamber.
\end{rem}

\begin{proof}
As the Weyl group permutes the Weyl chambers (see Fact~\ref{permWeyl}) 
we can choose
$a$ and $a'$ to be elements of the closed fundamental Weyl 
chamber. We recall from Eq.~(\ref{fundWeyl}) that an element 
$d=d_{1}X_{7}+d_{2}X_{8}+d_{3}X_{9}$
lies in the fundamental Weyl chamber iff
$d_{2}-d_{3}>0$, $d_{2}+d_{3}>0$, $d_{1}-d_{3}>0$, 
$d_{1}+d_{3}>0$, $d_{1}+d_{2}>0$, and $d_{1}-d_{2}>0$ holds.
Applying Thm.~\ref{thm iHsim2} and Fact~\ref{factweylchar} we get 
that
$a_{1} d_{1} + a_{2} d_{2} + a_{3} d_{3} \geq (a'_{1} d_{1} + 
a'_{2} d_{2} + a'_{3} d_{3})/t$ holds for all elements
$d=d_{1}X_{7}+d_{2}X_{8}+d_{3}X_{9}$ of the fundamental Weyl 
chamber.
Eliminating the quantifiers in the previous condition,
 e.g., using the computer algebra system
QEPCAD~\cite{CH:1991,QEPCADB}, we obtain 
the conditions of Eq.~(\ref{smajbed}).
\end{proof}

\subsection{Majorization\label{subsectmaj}}
In this subsection we introduce some concepts from the theory of
majorization which will be employed 
later.
Our presentation is succinct and we refer to
Refs.~\cite{NV:2001,MO:1979,AU:1982,And:1989,And:1994,Bha:1997} for a
more detailed treatment of this topic. 

For an element
$x=(x_{1},\ldots,x_{k})^{T}$ of $\R^{k}$   
we denote by
$x^{\downarrow}=(x^{\downarrow}_{1},\ldots,x^{\downarrow}_{k})^{T}$
a permutation of $x$ so that $x^{\downarrow}_{i}\geq 
x^{\downarrow}_{j}$ if $i<j$, where $1\leq i,j \leq k$.
\begin{defn}[Majorization {\cite[p.~28]{Bha:1997}}]
A vector $x \in \R^{k}$ is majorized by a vector $y \in \R^{k}$ if
\begin{gather*}
\sum^{l}_{i=1} x^{\downarrow}_{i} \leq \sum^{l}_{i=1} y^{\downarrow}_{i}
\text{ for all } 1\leq l \leq k\\
\intertext{and}
\sum^{k}_{i=1} x^{\downarrow}_{i} = \sum^{k}_{i=1}
y^{\downarrow}_{i}.
\end{gather*} 
The notation $x \prec y$ means that $x$ is majorized by $y$.
\end{defn}

We recall the notion of $s$-majorization
introduced in Ref.~\cite{BCL:02}. For an element
$x=(x_{1},x_{2},x_{3})^{T}$ of $\R^{3}$ we introduce the vector
$\hat{x}=(\lvert x_{1}\rvert,\lvert x_{2} \rvert,
\lvert x_{3} \rvert)^{T}$, and we define
the $s$-ordered version $x^{{\downarrow}_{s}}$  of $x$ by setting
$x^{{\downarrow}_{s}}_1 :=\hat{x}^{\downarrow}_{1}$,
$x^{{\downarrow}_{s}}_2 :=\hat{x}^{\downarrow}_{2}$,
and
$x^{{\downarrow}_{s}}_3 :=\sgn(x_{1} x_{2} x_{3})\hat{x}^{\downarrow}_{3}$.
The signum of $x_{1} x_{2} x_{3}$
is denoted by 
$\sgn(x_{1} x_{2} x_{3})$.
\begin{defn}[{\cite[p.~11]{BCL:02}}\label{defsmaj}]
The vector $x \in \R^{3}$ is $s$-majorized by $y \in \R^{3}$ if
\begin{align*}
x^{{\downarrow}_{s}}_{1} & \leq y^{{\downarrow}_{s}}_{1},\\
x^{{\downarrow}_{s}}_{1}  + x^{{\downarrow}_{s}}_{2} +
x^{{\downarrow}_{s}}_{3}
& \leq y^{{\downarrow}_{s}}_1 + y^{{\downarrow}_{s}}_{2}
+ y^{{\downarrow}_{s}}_{3},\\
x^{{\downarrow}_{s}}_{1}  + x^{{\downarrow}_{s}}_{2} -
x^{{\downarrow}_{s}}_{3}
& \leq y^{{\downarrow}_{s}}_1 + y^{{\downarrow}_{s}}_{2}
- y^{{\downarrow}_{s}}_{3}.
\end{align*}
The notation $x \prec_{s} y$ means that $x$ is $s$-majorized by $y$.
\end{defn}
We emphasize that a vector representing
an element from the Lie subalgebra $\ga$
is $s$-ordered if and only if
it lies in the closed fundamental Weyl chamber, as given
in Eq.~(\ref{fundWeyl}), except that for the closure
the relation $<$  has to be replaced by
the  relation $\leq$. This gives a geometric motivation for the $s$-ordered
vectors. In addition, the necessary and sufficient conditions 
for Hamiltonian simulation in Eq.~(\ref{smajbed}) are
equivalent to the definition of $s$-majorization.

\begin{cor}[{\cite[p.~11]{BCL:02}}\label{corHsimmaj}]
Assume that $H$ and $H'$ are non-local Hamilton operators
acting on a two-qubit system.
Let $a$ and $a'$ be  elements of $\ga$,
where
$a=a_{1}X_{7}+a_{2}X_{8}+a_{3}X_{9}=(\Ad_{\g}(L))(H)$
and 
$a'=a'_{1}X_{7}+a'_{2}X_{8}+a'_{3}X_{9}=(\Ad_{\g}(L'))(H')$
for some local unitary operators $L$ and $L'$.
We use the notation $\vec{a}=(a_{1},a_{2},a_{3})^{T}$ and
$\vec{a}'=(a'_{1},a'_{2},a'_{3})^{T}$.

A two-qubit system with Hamilton operator $H$ and local unitary
operators available is able to simulate the Hamilton operator $H'$ 
in time $t$ if and only if the following equation holds:
\begin{equation*}
\vec{a}' \prec_{s} t \vec{a}.
\end{equation*}
\end{cor}

There is a similar condition  on infinitesimal Hamiltonian 
simulation which is given  in terms of a majorization
condition on the spectra of the considered Hamilton operators.
This result (see appendix and Ref.~\cite[pp.~9--10]{VC:2002})
should be compared to the $s$-majorization condition 
in Cor.~\ref{corHsimmaj}.

\section{Gate simulation for two qubits}\label{gatesim}
As in Def.~\ref{defn gate simulation} we consider now 
gate simulation which is a global version of
infinitesimal Hamiltonian simulation. 
We recall a theorem of Khaneja et al.~\cite{KBG:2001}.

\begin{fact}[{\cite[Thm.~10]{KBG:2001}}\label{KBGthm}]
Assume that $H$ is a non-local Hamilton operator
acting on a two-qubit system.

A two-qubit system with Hamilton operator $H$ and local unitary
operators available is able to simulate the unitary gate $U$
in time $t$ if and only if the unitary gate $U$ can be decomposed
as
\begin{equation}\label{KBG Eq}
U = L_{1} \exp(t W) L_{2},
\end{equation}
where $L_{1}$ and $L_{2}$ are local unitary operators
and $W$ is an element which lies in the convex hull
of the Weyl orbit $\We(H)$ of $H$. 
\end{fact}

\begin{rem}
An equivalent version of Eq.~(\ref{KBG Eq}) is
\begin{equation}\label{KBG Eq2}
L^{-1}_{1} U L^{-1}_{2} = \exp(t W).
\end{equation}
This means that $U$ can be simulated in time $t$ if and only if
there exists a unitary gate $U'$ which is locally equivalent to $U$
and which can be expressed as $U' = \exp(t W)$.
But there exists a restriction on the elements $L_{1}$ and $L_{2}$.
As $\exp(t W)$ is an element of $\GA=\exp(\ga)$, we have
that $L^{-1}_{1} U L^{-1}_{2}$ has to be an element of
$\GA$, too. There exists different unitary operations $U'$
which satisfy this restriction.
The appearance of different unitary operations $U'$ is a 
consequence of the 
non-uniqueness of the $\GK\, \GA\, \GK$ decomposition of Fact~\ref{KAK}
which will be analyzed in detail below. 
We emphasize that it may be impossible to express $U$
as $U = \exp(t W)$ with the  same (or shorter) time~$t$
as in Eq.~(\ref{KBG Eq2}).
\end{rem}

We present now the results on gate simulation 
in similar fashion as done in Section~\ref{subsectmaj} for Hamiltonian 
simulation. Due to the remark following Fact~\ref{KBGthm}, a 
local unitary operation $U$ can 
be simulated in time $t$ if and only if a local 
unitary operation $U'$ which is locally
equivalent to $U$ can be expressed as $U'=\exp(tW)$, where $W$ denotes 
an element of the Weyl orbit of the system Hamiltonian. In the
sequel, let $K_{i}$, for 
$i \in \{1,\ldots, 8\}$, be suitable elements from the set of local
unitary gates $\GK=\exp(\gk)$. In addition we denote
by $A$ and $A'$ some appropriate elements of $\GA =\exp(\ga)$.
In view of Fact~\ref{KAK} we can decompose the unitary gate $U$
and locally equivalent gates $U'$ as $U=K_{1}AK_{2}$ and
$U'=K_{3}A'K_{4}$ respectively. To characterize all unitary gates
$U'$ which are locally equivalent to $U$ it is necessary and sufficient to
characterize all $A'$ satisfying
$K_{5}A'K_{6}=A$. Thus we have to identify all $A'$ which
can be written as $A'=(K_{7}^{-1}AK_{7})K_{8}$. This is done
in the following lemma.

\begin{lem}\label{KAKlem}
For a fixed $A \in \GA$ and an arbitrary element of the
form $A'=(K^{-1}AK)K' \in \GA$ with $K,K' \in \GK$ 
we can choose 
$K$ from the Weyl group and $K'$ from the set $\GK \cap \GA$.
\end{lem}

Related to $\theta$ from the definition
(Def.~\ref{osla}) of an orthogonal symmetric Lie algebra
$(\g,\theta)$, there exists a global version $\Theta$ operating
on the Lie group $\G$, see, e.g., 
Ref.~\cite[Thm.~2.3 of Chap.~IV]{Loo:1969a}
or Ref.~\cite[Thm.~6.31.]{Kna:2002}. We define $\Theta$ by
$\Theta(K'')=K''$ for $K'' \in \GK$ and $\Theta(P)=P^{-1}$ for 
$P \in \GP=\exp(\gp)$. We employ the mapping 
$()^{\star}\negthickspace : \G \to \G$
given by $G \mapsto G^{\star}:=\Theta(G^{-1})$. 
We use here the symbol $()^{\star}$ in order to avoid confusion
with the symbol $()^{*}$, which denotes complex conjugation.
We have that $(G_{1}G_{2})^{\star}=
G_{2}^{\star}G_{1}^{\star}$ for $G_{1},G_{2} \in \G$, $P^{\star}=P$ for 
$P \in \GP$, and $(K'')^{\star}=(K'')^{-1}$ for 
$K'' \in \GK$, see Ref.~\cite[p.~81]{Bor:1998}.
We introduce the map $\phi\negthickspace : \G/\GK \to \GP$ 
which is defined as
$GK \mapsto \phi(GK):=(GK)(GK)^{\star}=GG^{\star}$. This map $\phi$ was 
studied in Ref.~\cite[p.~81-82]{Bor:1998} and in 
Ref.~\cite[Proof of Thm.~6.31.]{Kna:2002}. Ref.~\cite{Bor:1998} proves
that $\phi$ induces an isomorphism
of $\G/\GK$ onto $\GP$.

\begin{proof}[Proof of Lemma~\ref{KAKlem}]
We employ the map $\phi$ and obtain the equations 
$\phi((K^{-1}AK)K')=K^{-1}A^{2}K$
and $\phi(A')=(A')^{2}$.
Since $A'=(K^{-1}AK)K'$ is given in the condition of Lemma~\ref{KAKlem},
we obtain $K^{-1}A^{2}K=(A')^{2}$.
And due to Fact~\ref{weylorbitlemma}, we can choose $K$ as an element of
the Weyl group. Thus, $K^{-1}AK \in \GA$ which proves that $K' \in \GK
\cap \GA$.
\end{proof}

We still need to characterize the elements of $\GK \cap \GA$.
This will be done now. 

\begin{lem}\label{KschnittA}
The elements of set $\GK \cap \GA$ are given by
$\exp(z_{1}\pi X_{7}+z_{2}\pi X_{8}+z_{3}\pi X_{9})$, where
$z_{j} \in {\Z}$ for $j \in \{1,2,3\}$ and $X_{7}$, $X_{8}$, $X_{9}$ 
as defined on page~\pageref{Xi}.
\end{lem}

\begin{proof}
First, we show that the elements $\exp(z_{1}\pi X_{7}+z_{2}\pi
X_{8}+z_{3}\pi X_{9})$ constitute a subset of $\GK \cap \GA$.
Since $\exp(z_{1}\pi X_{7}+z_{2}\pi X_{8}+z_{3}\pi X_{9})$ for
$z_{j} \in {\Z}$ are by definition elements of $\GA$
and $\GA$ is an Abelian group, we obtain that
\begin{align*}
&\exp(z_{1}\pi X_{7}+z_{2}\pi X_{8}+z_{3}\pi X_{9})\\
=&\exp(z_{1}\pi X_{7})\exp(z_{2}\pi X_{8})\exp(z_{3}\pi X_{9})\\
=&(\ii \sigma_{1}\otimes\sigma_{1})^{z_{1}}
(\ii \sigma_{2}\otimes\sigma_{2})^{z_{2}}
(\ii \sigma_{3}\otimes\sigma_{3})^{z_{3}}.
\end{align*}
This proves that the elements constitute a subset of $\GK \cap \GA$.
Secondly, we show that $\GK \cap \GA$ is a subset of 
the set given by the elements $\exp(z_{1}\pi X_{7}+z_{2}\pi
X_{8}+z_{3}\pi X_{9})$. We make the ansatz
$\exp(a_{7} X_{7}+a_{8}
X_{8}+a_{9} X_{9})=\exp(a_{1} X_{1}+a_{2}
X_{2}+a_{3} X_{3}+ a_{4} X_{4}+a_{5}
X_{5}+a_{6} X_{6})$, where $a_{i} \in {\R}$, $X_{i}$ were given
on p.~\pageref{Xi}, and $i \in \{1,\ldots,9\}$. By direct computations
one gets for $a_{7}$, $a_{8}$, and $a_{9}$ the conditions
\begin{gather*}
(a_{7}-a_{8}-a_{9})/\pi \in \Z, \\
(a_{7}+a_{8}-a_{9})/\pi \in \Z, \\
(a_{7}+a_{8}+a_{9})/\pi \in \Z, \\
(a_{7}-a_{8}+a_{9})/\pi \in \Z.
\end{gather*}
This implies $a_{i}/\pi \in \Z$ for $i \in \{7,8,9\}$.
\end{proof}
Now, we state the majorization-like
equivalent of Fact~\ref{KBGthm}.

\begin{cor}[see Ref.~{\cite[Lemma]{VHC:2002}} or 
Ref.~{\cite[Result~1]{HVC:2002}}\label{corKBG}]
Assume that $H$ is a non-local Hamilton operator
acting on a two-qubit system and that we intend to
simulate the unitary operation $U$.
Let $a$ and $a'$ be elements of $\ga$,
where $a=a_{1}X_{7}+a_{2}X_{8}+a_{3}X_{9}=(Ad_{\g}(K_{1}))(H)$,
$a'=a'_{1} X_{7} + a'_{2} X_{8} + a'_{3} X_{9}$,
and 
$U=K_{2} \exp(a') K_{3}$
for some local unitary operations $K_{1}, K_{2}, K_{3} \in \GK$.
We use the notation $\vec{a}=(a_{1},a_{2},a_{3})^{T}$
and $\vec{a}'=(a'_{1},a'_{2},a'_{3})^{T}$.

A two-qubit system with Hamilton operator $H$ and local unitary
operators available is able to simulate the unitary operation $U$
in time $t$ if and only if the following equation holds
for at least one choice of $\vec{z}=(z_{1},z_{2},z_{3})^{T} \in {\Z}^{3}$:
\begin{equation*}
\vec{a}' + \pi \vec{z} \prec_{s} t \vec{a}.
\end{equation*}
\end{cor}

\begin{proof}
By Fact~\ref{p char} and Fact~\ref{KAK} 
we can choose $a$ and $a'$ respectively as given.
Applying the remark following Fact~\ref{KBGthm} it is
necessary and sufficient to consider some unitary gates $U'$ which
are locally equivalent to $U$. 
By use of Fact~\ref{KAK} these locally equivalent gates $U'$
can be represented as $U'=K'_{1} A' K'_{2}$, where $A'$ is an element of $\GA$
and $K'_{1}, K'_{2}$ are local unitary gates. 
The different possibilities for $A'$ in this decomposition
are given by  Lemma~\ref{KAKlem} as
$A'=\exp\left[(Ad_{\g}(K))(a')\right] K'$,
where $K$ is an element of the Weyl group,
$K' \in \GK \cap \GA$, and $K'=\exp(k')$. 
With the characterization of $\GK \cap \GA$
from Lemma~\ref{KschnittA} we deduce that
$\GK \cap \GA$ is left invariant by
operations of the Weyl group.
Since $\GA$ is Abelian and $\GK \cap \GA$ is left invariant
by operations of the Weyl group, we can write
$A'$ as  $A'=\exp\left[(Ad_{\g}(K))(a')+k'\right]
=\exp\left[(Ad_{\g}(K))(a'+k'')\right]$,
where $K''=\exp(k'')$  for some element $K'' \in \GK \cap \GA$.
By Fact~\ref{KBGthm} we obtain that
$A'=\exp\left[(Ad_{\g}(K))(a'+k'')\right]=\exp(tW)$,
where $W$ lies in the convex hull of the
Weyl orbit $\We(a)$ of $a$.
When we consider the equation
$\exp\left[(Ad_{\g}(K))(a'+k'')\right]=\exp(tW)$
in a basis where both $(Ad_{\g}(K))(a'+k'')$
and $tW$ are diagonal then we obtain
by the periodicity of the exponential
function
that
$(Ad_{\g}(K))(a'+k'') + M =tW$, where 
$M=\diag(2\pi \ii \lambda_{1},2\pi \ii \lambda_{2},
2\pi \ii \lambda_{3},2\pi \ii \lambda_{4})$
and  $\lambda_{1},\lambda_{2},\lambda_{3},\lambda_{4} \in \Z$.
Since  $(Ad_{\g}(K))(a'+k'')$ and $tW$ are elements of $\ga$
it follows that $M \in \ga$.
We can write $M$ 
as $M=2\pi z_{1} X_{7}+2 \pi z_{2} X_{8}+2 \pi z_{3} X_{9}=2k_{1}$
where $z_{1}=(\lambda_{1}+\lambda_{2}) \in \Z$,
$z_{2}=(\lambda_{1}+\lambda_{3}) \in \Z$,
$z_{3}=(\lambda_{2}+\lambda_{3}) \in \Z$,
$K_{1}=\exp(k_{1})$, and  $K_{1} \in \GK \cap \GA$.
Thus, we obtain
$(Ad_{\g}(K))(a'+k'')+  2k_{1} =
(Ad_{\g}(K))(a'+k''+  2k_{2})=
(Ad_{\g}(K))(a'+k_{3})=tW$,
where $K_{i}=\exp(k_{i})$,  $K_{i} \in \GK \cap \GA$
and $i \in \{1,2,3\}$.
Corollary~\ref{corHsimmaj} completes the proof.
\end{proof}

Searching for a refinement of Cor.~\ref{corKBG}, we state bounds
on the coefficients of $a_{1}$, $a_{2}$, and $a_{3}$
of an element $a_{1}X_{7}+a_{2}X_{8}+a_{3}X_{9}$ of
$\ga$. 
It follows from
Lemma~\ref{KAKlem} and Lemma~\ref{KschnittA} that the 
coefficients $a_{i}$,
$i \in \{1,2,3\}$, are periodic with period $\pi$. 
(Concerning this periodicity, we refer also to 
Ref.~\cite[Appendix B]{KC:2001} and Ref.~\cite[p.~7]{ZVSW:2003}.)
Bearing the $\pi$-periodicity in mind, we can restrict
the coefficients $a_{i}$ to the interval $[-\frac{\pi}{2},\frac{\pi}{2}]$.
This choice is compatible with our conventions in 
Section~\ref{su4}.
To reduce the symmetry induced by the Weyl group, 
we restrict ourselves to elements
of the closed fundamental Weyl chamber. From Eq.~(\ref{fundWeyl}) or
from the $s$-order of Section~\ref{subsectmaj}, we get that
$a_{1} \geq 0$, $a_{2} \geq 0$, $a_{1} \geq a_{2}$, and 
$a_{2} \geq a_{3}$. These considerations lead to the following
corollary.

\begin{cor}[see Ref.~{\cite[Thm.~1]{VHC:2002}} or Ref.~{\cite[Result~2]{HVC:2002}}\label{corKBG2}]
Assume that $H$ is a non-local Hamilton operator
acting on a two-qubit system and that we intend to
simulate the unitary gate $U$.
Let $a$ and $a'$ be elements of $\ga$,
where $a=a_{1}X_{7}+a_{2}X_{8}+a_{3}X_{9}=(Ad_{\g}(K_{1}))(H)$,
$a'=a'_{1} X_{7} + a'_{2} X_{8} + a'_{3} X_{9}$,
and 
$U=K_{2} \exp(a') K_{3}$
for some local unitary operations $K_{1}, K_{2}, K_{3} \in \GK$.
In addition, we force $a_{1}$, $a_{2}$, $a_{3}$, $a'_{1}$, $a'_{2}$,
and $a'_{3}$ to be elements from the interval 
$[-\frac{\pi}{2},\frac{\pi}{2}]$.
We use the notation $\vec{a}=(a_{1},a_{2},a_{3})^{T}$
and $\vec{a}'=(a'_{1},a'_{2},a'_{3})^{T}$.

A two-qubit system with Hamilton operator $H$ and local unitary
operators available is able to simulate the unitary gate $U$
in time $t$ if and only if the following equation holds
for at least one choice of $\vec{z}=(z_{1},z_{2},z_{3})^{T} \in 
\{(0,0,0)^{T},(-1,0,0)^{T}\}$: 
\begin{equation*}
\vec{a}' + \pi \vec{z} \prec_{s} t \vec{a}.
\end{equation*}
\end{cor}

\begin{rem}
In the proof we follow Refs.~\cite{VHC:2002,HVC:2002}.
\end{rem}

\begin{proof}
Due to Cor.~\ref{corKBG} it is sufficient to proof that
for every $\vec{z} \in {\Z}^{3}$ one of the following conditions
holds:
\begin{align*}
\vec{a}' + \pi  (0,0,0)^{T} &\prec_{s} \vec{a}' + \pi \vec{z},\\
\vec{a}' + \pi  (-1,0,0)^{T}&\prec_{s} \vec{a}' + \pi \vec{z}.
\end{align*}
We first consider the case
that $\lvert z_{i} \rvert > 1$, for some $i \in \{1,2,3\}$.
Since 
$a'_{i} \leq \pi/2$, the maximal component 
$(\vec{a}' + \pi \vec{z})^{{\downarrow}_{s}}_{1}$
of the
$s$-ordered version of $\vec{a}' + \pi \vec{z}$ is greater than or
equal to
$2\pi - \pi/2=3 \pi/2$. We check the conditions of
Def.~\ref{defsmaj} and obtain that 
$\vec{a}' + \pi  (0,0,0)^{T} \prec_{s} \vec{a}' + \pi \vec{z}$.

Secondly, we consider the case
that $\lvert z_{i} \rvert \leq 1$ for all $i \in \{1,2,3\}$.
By easy, but tedious, computations one can check that
$\vec{a}' + \pi  (0,0,0)^{T} \prec_{s} \vec{a}' + \pi \vec{z}$
for
\begin{align*}
\vec{z} \in \{&(-1,-1,0)^{T},(-1,0,-1)^{T},(0,-1,-1)^{T},\\
&(0,-1,1)^{T},(0,0,0)^{T},(-1,0,1)^{T} \}
\end{align*}
and that $\vec{a}' + \pi  (-1,0,0)^{T} 
\prec_{s} \vec{a}' + \pi \vec{z}$ for 
\begin{align*}
\vec{z} \in 
\{&(-1,-1,-1)^{T},(-1,-1,1)^{T},(-1,0,0)^{T},\\
&(0,-1,0)^{T},(0,0,-1)^{T},(0,0,1)^{T} \}.
\end{align*}
For all other $\vec{z} \in \{-1,0,1\}^{3}$ we have that both
$\vec{a}' + \pi  (0,0,0)^{T} \prec_{s} \vec{a}' + \pi \vec{z}$
and $\vec{a}' + \pi  (-1,0,0)^{T} 
\prec_{s} \vec{a}' + \pi \vec{z}$ hold.
\end{proof}

\section{Lower bounds
for $n$-qubit systems\label{sectlow}}
In the two-qubit case we used a particular decomposition
$\g=\gk + \gp$ of the Lie algebra which leads to a decomposition 
$\G=\GK\, \GA\, \GK$ of the Lie group where $\GK=\exp(\gk)$
is the set of local unitary operations. 
By this approach, e.g., the optimal simulation result
of Fact~\ref{KBGthm} can be obtained.
In the more general $n$-qubit case 
we can use decompositions $\g=\gk + \gp$ of the corresponding Lie group 
where $\GK=\exp(\gk)$ contains all local unitary operations.
Although $\GK$ is in general not equal to the set of local unitary
operations we can generalize the approach from the two-qubit case
in order to prove lower bounds on the time complexity
for gate simulation. 
Lower bounds were considered in Ref.~\cite{CHN:2003}, and
we refine and generalize the approach of Ref.~\cite{CHN:2003}
in this section. In doing so, we put this approach in a broader context.

\subsection{Magic basis (for two qubits)\label{magic}}
We begin by recalling the Bell basis and
the magic basis. The Bell basis (see Refs.~\cite{BMR:1992,BBC:1993})
is a vector space basis for two-qubit
pure states:
\begin{align*}
|\Phi^{+}\rangle &:= 
\frac{1}{\sqrt{2}} \left( |00\rangle + |11\rangle \right),\\
|\Phi^{-}\rangle &:= 
\frac{1}{\sqrt{2}} \left( |00\rangle - |11\rangle \right),\\
|\Psi^{+}\rangle &:= 
\frac{1}{\sqrt{2}} \left( |01\rangle + |10\rangle \right),\\
|\Psi^{-}\rangle &:= 
\frac{1}{\sqrt{2}} \left( |01\rangle - |10\rangle \right).
\end{align*}
We employ the ket-vector notation, see, e.g., Ref.~\cite{NC:2000}.
If we include some relative phases in the Bell basis we get
the magic basis which was introduced in Ref.~\cite{BDS:1996} and
coined by Hill and Wootters \cite{HW:1997}: 
\begin{align*}
|e_{1}\rangle &:= |\Phi^{+}\rangle,\\
|e_{2}\rangle &:= \ii |\Phi^{-}\rangle,\\
|e_{3}\rangle &:= \ii |\Psi^{+}\rangle,\\
|e_{4}\rangle &:= |\Psi^{-}\rangle.
\end{align*}
The magic basis is connected to the entanglement of formation, see
Ref.~\cite{BDS:1996} and related work in
Refs.~\cite{BBP:1996,PR:1997}. We neglect here this
connection, but refer to Section~\ref{dis}.

The magic basis has two important properties. First, the local
unitary operations on two qubits are real and orthogonal in
the magic basis, see Ref.~\cite[p.~5023]{HW:1997} and Thm.~1 of
Refs.~\cite{Mak:2003}. Secondly, the elements of the 
$\GA=\exp(\ga)$ (for notations see, e.g., Sec.~\ref{su4}) are diagonal
in the magic basis, as remarked in Ref.~\cite[p.~3]{KC:2001} and
Ref.~\cite[p.~2]{HVC:2002}. The basis change from the
standard basis $\{|00\rangle,|01\rangle,|10\rangle,|11\rangle\}$
to the magic basis is given by $Q^{-1}$, where 
\begin{equation*}
Q =\frac{1}{\sqrt{2}}
\begin{pmatrix}
1 & 0 & 0 & \ii \\
0 & \ii & 1 & 0 \\
0 & \ii & -1 & 0 \\
1 & 0 & 0 & -\ii
\end{pmatrix}.
\end{equation*}
For elements $U \in \SU(4)$ the map $U \mapsto Q^{-1} U  Q$ 
(see Ref.~\cite{Mak:2003}) reflects 
the isomorphism between $\SU(2)\otimes\SU(2)$ and $\SO(4)$, see, e.g.,
Ref.~\cite[p.~52]{Gil:1994}.

\subsection{Representation theory\label{rep}}
It is not obvious how the magic basis generalizes to higher number
of qubits and which properties remain. 
Motivated by the properties of the 
magic basis for two qubits, we seek for basis changes
of the local unitary operations $(\SU(2))^{\otimes n}$ into the 
orthogonal group
(if possible). To analyze this we need some
representation theory.

\begin{defn}[Lie group representation, see, e.g., 
{Ref.~\cite[p.~210]{DK:2000}}]
A complex representation of the Lie group $\G$ in the
finite-dimensional and complex vector space $\V$ is a continuous 
homomorphism $\tau\negthickspace : \G \to \GL(\V)$ from the group $\G$
into the group $\GL(\V)$ of invertible and linear
transformations which operate on $\V$.
\end{defn}
A representation $\tau$ in a finite-dimensional and complex vector
space $\V$ is called irreducible if there exists no subspace
$\UC$ other than $\UC={\mathrm 0}$ or $\UC=\V$ such that
the subspace is $\tau(\G)$-invariant, i.e.,
the equation $\tau(\G)\UC \subset \UC$ holds (see, e.g., 
Ref.~\cite[p.~210]{DK:2000}).
We state an important fact on tensor products of irreducible
representations.
\begin{fact}[{\cite[Prop.~4.14 of Chap.~II]{BD:1985}}]
If $\tau_{1}$ is an irreducible complex representation of ${\G}_{1}$ 
in the complex vector space $\V$
and $\tau_{2}$ is an irreducible complex representation of ${\G}_{2}$
in the complex vector space $\W$,
then $\tau_{1}\otimes\tau_{2}$ is an irreducible complex representation
of ${{\G}_{1}}\times {{\G}_{2}}$
in the complex vector space $\V \otimes \W$. Furthermore,
any irreducible representation of ${{\G}_{1}}\times {{\G}_{2}}$ is a tensor
product of this form.
\end{fact}
Below we use bilinear forms $\B\negthickspace : \V \times \V \to \C$, 
which are 
$\C$-linear in both arguments, to characterize irreducible complex
representations. Let $v_{1}$ and $v_{2}$ be some
arbitrary elements of $\V$. A bilinear form is called symmetric if
$\B(v_{1},v_{2})=\B(v_{2},v_{1})$ and skew-symmetric if 
$\B(v_{1},v_{2})=-\B(v_{2},v_{1})$. 
A bilinear form is $\tau(\G)$-invariant if 
$\B(v_{1},v_{2})=\B(\tau(G)v_{1},\tau(G)v_{2})$ for all
$G \in \G$.

\begin{defn}[cf. {Refs.~\cite{BD:1985,DK:2000}}\label{reptype}]
Consider an irreducible complex representation $\tau$
of $\G$ in $\V$.
The representation $\tau$ is said to be of 
\begin{itemize}
\item real type if $\V$ admits a bilinear form which
is nonzero, non-degenerate, $\tau(\G)$-invariant, and symmetric,
\item complex type if $\V$ admits no bilinear form which
is nonzero, non-degenerate, and $\tau(\G)$-invariant,
\item quaternionic type if $\V$ admits a bilinear form which
is nonzero, non-degenerate, $\tau(\G)$-invariant, and skew-symmetric.
\end{itemize}
\end{defn}

We introduce the map 
$\chi_{\tau}\negthickspace : \G \to \C, G \mapsto \Tr(\tau(G))$
which is the character $\chi_{\tau}$ of the representation $\tau$. We
use the character to characterize the type (real, complex, or
quaternionic) of irreducible complex representations.

\begin{fact}[{\cite[Thm.~4.8.1]{DK:2000}}\label{reptype1}]
Let $\tau$ denote an irreducible complex representation of $\G$ in $\V$.
The character $\chi_{\tau}$ is real-valued if and only if
there exists a $\tau(\G)$-invariant, nonzero, complex bilinear form
$\B$ on $\V$ which is automatically non-degenerate and uniquely
determined, up to a nonzero scalar factor. This bilinear form
$\B$ is either symmetric or skew-symmetric.
\end{fact}

By means of Fact~\ref{reptype1} we can decide if the type of a
representation is complex. To complete the classification of the 
type (real, complex, or quaternionic) of irreducible complex 
representations, we state another fact which allows to determine
the type of a representation by computing an normalized
integral over the compact Lie group~$\G$.

\begin{fact}[see {Ref.~\cite[Prop.~4.8.7]{DK:2000}} and
{Ref.~\cite[Prop.~6.8 of Chap.~II]{BD:1985}}\label{chi_integral}]
Let $\tau$ be an irreducible complex representation of the compact Lie
group $\G$ in $\V$
with character $\chi_{\tau}$. Then we have
\begin{equation*}
\int \chi_{\tau}(G^2) d\G = 
\begin{cases}
1& \Leftrightarrow \text{ $\tau$ is of real type,}\\
0& \Leftrightarrow \text{ $\tau$ is of complex type,}\\
-1& \Leftrightarrow \text{ $\tau$ is of quaternionic type.}
\end{cases}
\end{equation*}
\end{fact}

Representations can be identified with subgroups of $\GL(\V)$, so we
can extend Def.~\ref{reptype} to subgroups of $\GL(\V)$. We denote
the general linear group on a complex vector space of dimension
$k$ by $\GL(k,\C)$. Next we characterize the subgroups of $\GL(k,\C)$
that are conjugated to subgroups of the orthogonal
group  (motivated by Sec.~\ref{magic}) or the symplectic group.

\begin{fact}[adapted from {Ref.~\cite[Thm.~H of Chap.~3]{Sam:1990}}]
A compact subgroup of the general linear group $\GL(k,\C)$ is
conjugated in $\GL(k,\C)$ to a subgroup of the orthogonal
group $\Or(k)$ if and only if it is of real type. Accordingly, a subgroup
of $\GL(2k,\C)$ is conjugated in $\GL(2k,\C)$ to a subgroup of the
(unitary) symplectic group $\SP(k)$ if and only if it is
of quaternionic type.
\end{fact}

\begin{rem}
Actually, Ref.~\cite{Sam:1990} gives an algorithm to compute
the basis change from the bilinear form mentioned in
Def.~\ref{reptype}. For the notation
$\SP(k)$ see Subsection~\ref{lowbounds} and Ref.~\cite{Che:1999}.
\end{rem}

After this preparation we consider the case of local unitary operations
$(\SU(2))^{\otimes n}$. We employ the standard representation
of $\SU(2)$:
\begin{equation*}
G=
\begin{pmatrix}
a + \ii b & c + \ii d\\
-c+ \ii d & a - \ii b
\end{pmatrix},
\end{equation*}
where $a,b,c,d \in \R$ and $a^2+b^2+c^2+d^2=1$. 
To compute the integral of
Fact~\ref{chi_integral}, we introduce the
real parameters $0 \leq \phi < 2 \pi$, $0 \leq \psi_{1} \leq \pi$, and
$0 \leq \psi_{2} \leq \pi$ as follows:
\begin{align*}
a &=\cos(\phi) \sin(\psi_{1}) \sin(\psi_{2}),\\
b &=\sin(\phi) \sin(\psi_{1}) \sin(\psi_{2}),\\
c &=\cos(\psi_{1}) \sin(\psi_{2}),\\
d &=\cos(\psi_{2}).
\end{align*}
We obtain
\begin{align*}
&\int \chi_{\tau}(G^2) d (\SU(2))^{\otimes n}\\
&=\left(\int \chi_{\tau}(G^2) d \SU(2)\right)^{n}\\
&=\left(\int\limits_{0}^{2\pi} \int\limits_{0}^{\pi}
\int\limits_{0}^{\pi} \Xi(\phi,\psi_{1},\psi_{2})
d\psi_{2}d\psi_{1}d\phi\right)^{n}\\
&=(-1)^{n},
\end{align*}
where
\begin{align*}
\Xi(\phi,\psi_{1},\psi_{2}) =
&[  4 \cos(\phi)^{2}
( 1 - \cos(\psi_{1})^{2} - \cos(\psi_{2})^{2} \\
& +  \cos(\psi_{1})^{2} \cos(\psi_{2})^{2} 
) - 2 ] \sin(\psi_{1}) \sin(\psi_{2})^{2}.
\end{align*}
This proves the following theorem.
\begin{thm}\label{orthsymp}
The local unitary operations on an even number of qubits are
conjugated to a subgroup of an orthogonal group.
The local unitary operations on an odd number of qubits are
conjugated to a subgroup of a (unitary) symplectic group.
\end{thm}

\begin{rem}
Similar results as in this subsection are obtained in
Ref.~\cite{BB:2003} using a different approach.
\end{rem}

\subsection{Thompson's theorem and majorization}
Following Ref.~\cite{CHN:2003}, we present 
in this subsection a theorem due to Thompson \cite{Tho:1986}
and a majorization condition for the spectra of the sum of two Hermitian
matrices. Both results will be employed below.
\begin{fact}[\cite{Tho:1986}]
Let $A$ and $B$ be Hermitian matrices. Then there exist 
unitary matrices $U_{1}$ and $U_{2}$ such that
\begin{equation*}
\exp(\ii A) \exp(\ii B) = 
\exp(\ii U_{1}^{-1} A U_{1} + \ii  U_{2}^{-1} B U_{2} ). 
\end{equation*} 
\end{fact}
This result of  Thompson relies partly on a conjecture of Horn
\cite{Hor:1962}. This conjecture was recently proven
\cite{Lid:1982,DST:1998,Kly:1998,KT:1999,Knu:2000,Ful:2000}.
By induction, we get the following corollary.
\begin{cor}\label{thompsoncor}
Let $A_{j}$ denote Hermitian matrices. Then there exist
unitary matrices $U_{j}$ such that
\begin{equation*}
\prod_{j=1}^{m} \exp(\ii A_{j}) = \exp\left(\ii \sum_{j=1}^{m} U_{j}^{-1} A_{j} U_{j}\right).
\end{equation*}
\end{cor}

We state now a result which gives us bounds for the
the spectra of the sum of two Hermitian matrices. Reference~\cite{MO:1979}
attributes this result to Ky Fan \cite{Fan:1949}.
We denote the vector of eigenvalues of the $k\times k$-dimensional matrix 
$A$, including multiplicities, by $\spec(A)
=(\spec(A)_{1},\ldots,\spec(A)_{k})^{T}$. In addition, we assume
that $\spec(A)_{i}\geq \spec(A)_{j}$ if $i<j$
($1\leq i,j \leq k$).

\begin{fact}[{\cite[Thm.~9.G.1.]{MO:1979}}\label{specmaj}]
Let $A$ and $B$ denote Hermitian matrices.
Then the following equation holds:
\begin{equation*}
\spec(A+B) \prec \spec(A)+\spec(B).
\end{equation*}
\end{fact}

\subsection{Lower bounds\label{lowbounds}}
In this subsection we derive lower bounds on the minimal time 
to simulate unitary operations (see Def.~\ref{defn gate simulation}). 
We begin by discussing the (unitary)
symplectic group. Following Ref.~\cite[p.~22]{Che:1999}, we 
introduce the bilinear form
\begin{equation}\label{bilinear}
\B_{\SP}(\vec{x},\vec{y}):=
\sum_{j=1}^{k} (x_{i}y_{i+k}-x_{i+k}y_{i}),
\end{equation}
where 
$\vec{x}=(x_{1},\ldots,x_{2k})^{T} \in \C^{2k}$ and
$\vec{y}=(y_{1},\ldots,y_{2k})^{T} \in \C^{2k}$.
Let $J_{k}$ denote the matrix 
\begin{equation*}
J_{k}=
\begin{pmatrix}
0_{k} & I_{k}\\
-I_{k} & 0_{k}
\end{pmatrix},
\end{equation*}
where $I_{k}$ is the $k\times k$-dimensional identity matrix
and $0_{k}$ the $k\times k$-dimensional zero matrix.

\begin{defn}[see, e.g., {Ref.~\cite[Prop.~1 on p.~22]{Che:1999}}]
The subgroup of the unitary group $\U(2k)$ of degree $2k$
composed of the matrices $M$ which leave the bilinear form 
in Eq.~(\ref{bilinear}) invariant, i.e., which satisfy the condition 
\begin{equation}\label{eqJ}
M^{T} J_{k}  M = J_{k},
\end{equation}
is called the (unitary) symplectic group and is 
denoted by $\SP(k)$.
\end{defn}

The group $\SP(k)$ can be considered as operating on a $k$-dimensional
module over the quaternions $\Q$ leaving  a symplectic (scalar)
product invariant \cite[pp.~16--24]{Che:1999}. All elements of $\SP(k)$
have determinant one, see, e.g., Ref.~\cite[p.~203]{Che:1999}.
When we regard $\SP(k)$ as a manifold its
real dimension is $2k^2+k$ \cite[p.~23]{Che:1999}.

We recall from Thm.~\ref{orthsymp} that the local unitary operations
on an odd number of qubits are conjugated to a subgroup of a (unitary) 
symplectic group. 
Using that $(J_{k})^{-1}=-J_{k}$, the
condition in Eq.~(\ref{eqJ}) can be proved to be equivalent to
\begin{equation}\label{eqJ2}
M^{-1}=J_{k} M^{T} (J_{k})^{-1}.
\end{equation}
We know that the local unitary operations on an
odd number of qubits meet the condition in Eq.~(\ref{eqJ2})
in some appropriate chosen basis. 
But we can state the condition
also in the standard representation of $\SU(2)^{\otimes n}$ with $n$ odd.
We use the identification  $2k=2^n$.
Let $J_{n}'$ denote the matrix
\begin{equation*}
J_{n}':=
{\begin{pmatrix}
0 & 1\\
-1 & 0
\end{pmatrix}}^{\otimes n}
=
(\ii \sigma_{y})^{\otimes n}
\end{equation*}
and recall the standard representation
of $\SU(2)$:
\begin{equation*}
G=
\begin{pmatrix}
a + \ii b & c + \ii d\\
-c+ \ii d & a - \ii b
\end{pmatrix},
\end{equation*}
where $a,b,c,d \in \R$.
We use the notation
\begin{equation*}
G_{j}=
\begin{pmatrix}
a_{j} + \ii b_{j} & c_{j} + \ii d_{j}\\
-c_{j} + \ii d_{j} & a_{j} - \ii b_{j}
\end{pmatrix},
\end{equation*}
where $a_{j},b_{j},c_{j},d_{j} \in \R$.
It can be checked that
\begin{equation*}
{\begin{pmatrix}
0 & 1\\
-1 & 0
\end{pmatrix}}
G^{T}
{\begin{pmatrix}
0 & 1\\
-1 & 0
\end{pmatrix}}^{-1}
=
G^{-1}
\end{equation*}
and we obtain that 
\begin{equation}\label{tildeinverse}
J_{n}'
\left(
\bigotimes_{j=1}^{n}
G_{j}
\right)^{T}
(J_{n}')^{-1}
=
\left(
\bigotimes_{j=1}^{n}
G_{j}
\right)^{-1},
\end{equation}
which holds obviously for $n$ odd and even. 
From now on, $n$ is no longer restricted
to be odd. We emphasize
that $(J_{n}')^{-1}=(J_{n}')^{T}=(-1)^{n} J_{n}'$.
It follows that
\begin{equation}\label{Ginv}
\left(
\bigotimes_{j=1}^{n}
G_{j}
\right)^{T}
J_{n}'
\left(
\bigotimes_{j=1}^{n}
G_{j}
\right)
=
J_{n}'.
\end{equation}

Let $\HV$ denote the 
$2^n$-dimensional
complex vector space 
on which the group $\SU(2^n)$ operates.
We introduce the bilinear form $\B_{\HV}(x,y):=x^{T}J_{n}'y$
on the Hilbert space $\HV$. 
We have that
\begin{gather*}
\B_{\HV}(y,x)=
y^T J_{n}' x 
=
(-1)^{n}
x^T J_{n}' y
=
(-1)^{n} \B_{\HV}(x,y),
\end{gather*}
which proves that $\B_{\HV}(x,y)$ is symmetric for $n$ even
and skew-symmetric for $n$ odd. 
From Eq.~(\ref{Ginv}) we get that $\B_{\HV}(x,y)$
is left invariant by $\SU(2)^{\otimes n}$.
Hence, we have identified 
$\B_{\HV}(x,y)$ as the bilinear form of Def.~\ref{reptype}
operating on  $\V =\HV$.

This motivates the following definition of the tilde mapping,
which operates on the local unitary operations as the inverse
operation.
\begin{defn}\label{tilde}
We introduce the tilde mapping
\begin{equation*}
\Psi:
\left\{
\begin{aligned}
&\SU(2^n) \to \SU(2^n)\\
&U  \mapsto \tilde{U} := \Psi(U) = J_{n}'
U^{T}
(J_{n}')^{-1}
\end{aligned}
\right. 
\end{equation*}
\end{defn}

It is apparent from 
$
[J_{n}' U^{T}(J_{n}')^{-1}]
[J_{n}' U^{T}(J_{n}')^{-1}]^{\dagger}
=I_{2^n}$
and
$
\det[(J_{n}')^{-1} U^{T} J_{n}']
=\det(U)
$ that the tilde mapping preserves the  group $\SU(2^n)$.

\begin{rem}
The tilde mapping is a generalization
of the map $U \mapsto (\sigma_{y})^{\otimes n} U^{T} 
(\sigma_{y})^{\otimes n} $ for even $n$ 
from Ref.~\cite[p.~5]{CHN:2003}. It can be easily checked
that the two maps coincide in the case of even $n$.
See also the discussion in Sec.~\ref{dis}.
\end{rem}

We state now an important lemma characterizing the tilde
mapping.

\begin{lem}
Let $V$ and $W$ denote some local unitary operations and
let $U$ denote some arbitrary unitary operation. The following
equations hold:
\begin{enumerate}
\item $\tilde{V}=V^{-1}$,
\item $\tilde{W}=W^{-1}$,
\item $VUW\Psi{(VUW)}=VU\tilde{U}V^{-1}$.
\end{enumerate}
\end{lem}

\begin{proof}
The first and second claim follows from Eq.~(\ref{tildeinverse}).
We prove now the third claim:
$
VUW\Psi{(VUW)}=
VUW\tilde{W}\tilde{U}\tilde{V}
=VUWW^{-1}\tilde{U}V^{-1} =
VU\tilde{U}V^{-1}.
$
\end{proof}

This proves that local
unitary operations  preserve 
the spectrum of $U\tilde{U}$. 
We state now the theorem which gives us lower 
bounds for the  minimal time
to simulate a unitary gate.
We use the notation $\arg$ where
$\arg\left[\exp(\ii m)\right]=m$ and
$\arg\left[(x_1,\ldots,x_l)^T\right]=
(\arg[x_1],\ldots,\arg[x_l])^T$.

\begin{thm}
Assume that $H$ is a non-local Hamilton operator
acting on an $n$-qubit system and that
we intend to simulate the unitary gate $U$.
An $n$-qubit system with Hamilton operator $H$ and
local unitary operators available is able to simulate the
unitary gate $U$ in time $t$ only if
the following equation holds for at least
one choice of $\vec{z}\in {\Z}^{2^n}$:
\begin{equation*}
\arg\left[\spec(U\tilde{U})\right]+2\pi \vec{z} \prec 2t\,  \spec(H)  
\end{equation*}
\end{thm}

\begin{rem}
This theorem generalizes the work in Ref.~\cite[Thm.~5 and Cor.~7]{CHN:2003}.
In the proof we use ideas from Ref.~\cite{CHN:2003}.
\end{rem}

\begin{proof}
Assume that $U=[ \prod_{j=1}^{m} W_{j} \exp(- \ii t_{j} H) ] W_{0}$
is a simulation of $U$,
where $W_{j}$ denotes some local unitary operations,
$t_{j}\geq 0$, and $\sum_{j=1}^{m} t_{j} = t$.
Let 
\begin{equation*}
{V}_{j} = 
\begin{cases}
W_{m}& \text{for $j=m$},\\ 
{V}_{j+1}  W_{j}& \text{for $0\leq j <m$}.
\end{cases}
\end{equation*}
and $H_{j}=V_{j} H V_{j}^{-1}$ ($1\leq j \leq m$).
Thus, the simulation of $U$ can
written as $U = [ \prod_{j=1}^{m}  \exp(- \ii t_{j} H_{j}) ] V_{0}$
where $\spec(H_{j})=\spec(H)$ for all $1\leq j \leq m$.

We use Cor.~\ref{thompsoncor} to find some
suitable Hermitian operators $H_{j}'$ and $H_{j}''$
with $\spec(H_{j}')=\spec(H_{j}'')=\spec(H)$ and 
$j \in \{1,\ldots,m\}$
such that
\begin{eqnarray}\label{hilfseq}
\left[\prod_{j=1}^{m} \exp(- \ii t_{j} H_{j})\right] 
\Psi{\left(\prod_{j=1}^{m} \exp(- \ii t_{j} H_{j})
\right)}\nonumber \\
=\exp\left[\sum_{j=1}^{m} - \ii t_{j}(H_{j}' + H_{j}'')\right].
\end{eqnarray}

When we combine $\tilde{V_{0}}=V_{0}^{-1}$ with Eq.~(\ref{hilfseq})
we obtain that
\begin{equation*}
\spec(U\tilde{U})=
\spec\left\{\exp\left[\ii \sum_{j=1}^{m} t_{j} 
(H_{j}'+H_{j}'') \right]\right\}.
\end{equation*}
We employ Fact~\ref{specmaj} to complete the proof:
\begin{align*}
\arg\left[\spec(U\tilde{U})\right] + 2 \pi \vec{z}
&=\spec\left[\sum_{j=1}^{m} t_{j} (H_{j}'+H_{j}'') \right]\\
&\prec 2t\, \spec(H).
\end{align*}
\end{proof}

\subsection{Involutive automorphisms}
We end this section by highlighting connections
between the tilde mapping of Def.~\ref{tilde} and involutive automorphisms
of the Lie algebra $\su(2^n)$.

The tilde mapping is similar to the $()^{\star}$-map used in the
proof of Lemma~\ref{KAKlem} in Section~\ref{gatesim}.
For $n$ odd $\GK$ must be equivalent to $\SP(2^{n-1})$ 
and respectively for $n$ even $\GK$ must  be equivalent to $\SO(2^n)$.
In both cases, the map $U \mapsto U\tilde{U}$ plays a similar r\^ole
as the map $\phi$ in Section~\ref{gatesim}.

Following Ref.~\cite[pp
.~451--452]{Hel:2001} we state all 
Riemannian symmetric spaces $\SU(2^n)/K$ induced
by involutive automorphisms of the Lie algebra $\su(2^n)$.
We have to consider three cases which correspond to
the types AI, AII, and AIII of involutive automorphisms.
In the case of type AI we have to treat the Lie algebra
$\g = \su(k)$ and the involutive automorphism $\theta_{\text{AI}}(g)=g^{*}$.
The involutive automorphism $\theta_{\text{AI}}$ gives rise to
the Riemannian symmetric space $\SU(k)/\SO(k)$.

The Lie algebra $\g = \su(2k)$ and the involutive automorphism
$\theta_{\text{AII}}(g)=J_{k} g^{*} (J_{k})^{-1}$ belong to type AII.
We obtain the Riemannian symmetric space $\SU(2k)/\SP(k)$.

For completeness we mention the type AIII even though
we do not use the corresponding Riemannian symmetric space
in this paper. The Lie algebra is $\g=\su(p+q)$ and the corresponding
involutive automorphism is given by
$\theta_{\text{AII}}(g)=I_{p,q} g I_{p,q}$. We have used
the notation 
\begin{equation*}
I_{p,q}=
\begin{pmatrix}
-I_{p} & 0_{p,q}\\
0_{q,p} & I_{q}
\end{pmatrix},
\end{equation*}
where $I_p$ denotes the $p\times p$-dimensional identity 
matrix and $0_{p,q}$ denotes the $p\times q$-dimensional
zero matrix.
This gives us the Riemannian symmetric space
$\SU(p+q)/{\mathrm{S}}(\U(p)\times \U(q))$.
The group ${\mathrm{S}}(\U(p)\times \U(q))$ can be represented by
the matrices
\begin{equation*}
\begin{pmatrix}
g_{1} &0_{p,q}\\
0_{q,p} &g_{2}
\end{pmatrix},
\end{equation*}
where $g_{1} \in \U(p)$,  $g_{2} \in \U(q)$, and
$\det(g_{1})\det(g_{2})=1$.

\section{Related Work\label{relwork}}
To recognize the considerable amount of related  work 
we give a short outline of the connections to 
our work. Various aspects of
(infinitesimal) Hamiltonian simulation as considered
in Sect.~\ref{infini} were studied in 
Refs.~\cite{VLK:1999,JK:1999,LCYY:2000,DVC:2001,SM:2001,DNBT:2002,WJB:02,WJB:02b,BCL:02,Leung:2002,VC:2002,VC:2002b,WRJB:02,NBD:2002,WRJB:02b,MVL:02,Chen:2003,CLV:2003,HNO:2003}.
Some references consider models where---in contrast to our 
model---additional resources were used: prior
entanglement~\cite{VC:2002b}, additional classical
communication~\cite{VC:2002}, measurements~\cite{BCL:02}, or
ancillas~\cite{BCL:02,VC:2002}. 

For two qubits, gate simulation (see Sect.~\ref{gatesim}) 
was analyzed in 
Refs.~\cite{KBG:2001,VHC:2002,HVC:2002,BDD:2002,ZVSW:2003,HNO:2003}. 
In Ref.~\cite{KG:2001b} Lie group
decompositions were used to obtain
a theory of $n$-qubit gate simulation. 
In general these decompositions do not lead to optimal simulations.
In the case of three qubits some 
progress on the time optimality problem for gate simulation 
was reported in Ref.~\cite{KGB:2002}, see also Ref.~\cite{KGB:2003}.
Concerning lower bounds,
we have generalized (see Sect.~\ref{sectlow}) the approach of
Ref.~\cite{CHN:2003}.

\section{Discussion\label{dis}}
In this section we address  a peculiar  similarity
between our approach to lower bounds 
on the time complexity for gate simulation
and 
the concurrence \cite{HW:1997,Woo:1998,Woo:1998b,Woo:2001},
as well as some of its generalizations 
\cite{CKW:2000,Uhl:2000,WC:2001,AF:2001,RBC:2001,AVM:2001,BDH:2002,JTS:2003,FMI:2003,LBZ:2003,TJ:2003,Ger:2003,VDM:2003,JSS:2003,BB:2003}.
The concurrence $C$ of a pure two-qubit state $|\psi\rangle \in
{\C}^{4}$  was 
defined in Ref.~\cite{Woo:1998} as
\begin{equation*}
C(|\psi\rangle)=\lvert\langle\psi|\tilde{\psi}\rangle\rvert,
\end{equation*}
where $|\tilde{\psi}\rangle := 
(\sigma_{y} \otimes \sigma_{y}) (|\psi\rangle)^{*}$. 
Let $\lambda_{1}$, $\lambda_{2}$, $\lambda_{3}$, and $\lambda_{4}$
denote the (positive) square roots of the
eigenvalues of the matrix $\rho \tilde{\rho}$, where 
$\tilde{\rho}:=(\sigma_{y} \otimes \sigma_{y}) \rho^{*} 
(\sigma_{y} \otimes \sigma_{y})$. We assume that
$\lambda_{1} \geq \lambda_{2} \geq \lambda_{3} \geq \lambda_{4}$.
References~\cite{HW:1997,Woo:1998} show that the concurrence $C$
of a two-qubit density matrix $\rho$ is given by
\begin{equation}\label{concurrence}
C(\rho)=\max\{0,\lambda_{1}-\lambda_{2}-\lambda_{3}-\lambda_{4}\}.
\end{equation}

Uhlmann \cite{Uhl:2000} considered generalizations of the
concurrence. Following this approach we introduce some
notations.
Let us call a map $\vartheta$ that operates
on
a complex vector space $\V$ antilinear if the equation
$
\vartheta(b_{1} |\psi_{1}\rangle + b_{2} |\psi_{2}\rangle )=
(b_{1})^{*} \vartheta(|\psi_{1}\rangle) + 
(b_{2})^{*} \vartheta(|\psi_{2}\rangle)
$
holds for all $b_{1},b_{2}
\in \C$ and all $|\psi_{1}\rangle,|\psi_{2}\rangle \in \V$.
For an antilinear operator  $\vartheta$ the (Hermitian) adjoint
$\vartheta^{\dagger}$ is defined by the condition that
$\langle\psi_{1}|\vartheta^{\dagger}\psi_{2}\rangle
= \langle\psi_{2}|{\vartheta}\psi_{1}\rangle$ holds
for all $|\psi_{1}\rangle,|\psi_{2}\rangle \in \V$.
If an antilinear operator $\vartheta$ satisfies the condition
${\vartheta}^{\dagger}={\vartheta}^{-1}$ we call
this operator  antiunitary.
When the map $\vartheta$ is antiunitary and 
${\vartheta}^{-1}=\vartheta$ holds, then we have that ${\vartheta}^{2}$
equals the identity map  and we define $\vartheta$ to be a conjugation.
A skew conjugation is an antiunitary operator $\vartheta$
fulfilling ${\vartheta}^{-1}=-\vartheta$.
Assume in the following that $\vartheta$ is a conjugation.
Now, Uhlmann defined a generalized tilde mapping by its operation on 
pure states $|\tilde{\psi}\rangle:=\vartheta(|\psi\rangle)$ and 
its operation 
on density matrices 
$\tilde{\rho}:=\vartheta \rho {\vartheta}^{-1} =\vartheta \rho \vartheta$.
In addition he generalizes the concept of concurrence 
to more than two qubits 
for pure states
\begin{equation*}
C_{\vartheta}(|\psi\rangle):=\lvert \langle\psi|\tilde{\psi}\rangle \rvert 
\end{equation*}
and for mixed states
\begin{equation*}
C_{\vartheta}(\rho):=\min \sum_{j} \lvert\langle\phi_{j}|\tilde{\phi}_{j}
\rangle\rvert,  
\end{equation*}
where the minimum is taken over all decompositions 
$
\rho = \sum_{j} |\phi_{j}\rangle\langle\phi_{j}|
$
of $\rho$ into non-normalized pure states $|\phi_{j}\rangle$.
Uhlmann \cite{Uhl:2000} proved that in strong analogy to 
Eq.~(\ref{concurrence})
the generalized concurrence is given by
\begin{equation*}
C_{\vartheta}(\rho)=\max\left\{0,\lambda_{1}-\sum_{j>1}\lambda_{j}\right\},
\end{equation*}
where the $\lambda_{i}$s are the square roots of the eigenvalues
of the matrix $\rho \tilde{\rho}$ and $\lambda_{i} \geq \lambda_{j}$
for $i<j$.

References~\cite{JTS:2003,TJ:2003,JSS:2003}
consider the map on density matrices given by
\begin{equation}\label{rho}
\rho \mapsto \tilde{\rho}:=(\sigma_{y})^{\otimes n} \rho^{*} 
(\sigma_{y})^{\otimes n}.
\end{equation}
In addition, the
map 
\begin{equation}\label{iH}
\ii H \mapsto \widetilde{\ii H}:=(- \ii \sigma_{y})^{\otimes n} (\ii H)^{*} 
\left( (- \ii \sigma_{y})^{\otimes n} \right)^{-1}
\end{equation}
is introduced in Ref.~\cite{BB:2003} for elements $\ii H$ of
the Lie algebra $\su(2^n)$. The map in Eq.~(\ref{iH}) can 
be applied to a Hamilton operator $H$:
\begin{equation*}
\tilde{H} = (- \ii \sigma_{y})^{\otimes n} (H)^{*} 
\left( (- \ii \sigma_{y})^{\otimes n} \right)^{-1}
= ( \sigma_{y})^{\otimes n} (H)^{*} 
 (  \sigma_{y})^{\otimes n}.
\end{equation*}
This shows that both Eq.~(\ref{rho}) and Eq.~(\ref{iH})
are induced by the conjugation $\vartheta_{1}$ given by
\begin{equation*}
\vartheta_{1}(|\psi\rangle)=\sigma_{y}^{\otimes n} (|\psi\rangle)^{*}.
\end{equation*} 
In this case we get the concurrence $C_{\vartheta_{1}}$.
The corresponding tilde mapping is given by its
action on pure states $|\tilde{\psi}\rangle=\vartheta_{1}|\psi\rangle$ 
and its action on density matrices
$\tilde{\rho}=\vartheta_{1}\rho {\vartheta_{1}}^{-1}=
\vartheta_{1}\rho \vartheta_{1}$.
A result of Ref.~\cite[Prop.~8]{VW:2000} (for related remarks see 
Ref.~\cite{RBC:2001}) states that the conjugation $\vartheta_{1}$ is 
the (up to a phase) unique antilinear operator acting on the complex vector 
space $(\C^2)^{\otimes n}$ which is invariant under basis 
changes by local unitary operations $U$ except for an factor 
equal to $\det(U)$. Further on, Ref.~\cite{VW:2000} states that such 
an antilinear mapping exists only for $n$-qubit systems, and not
for general $n$-qudit systems.

After this short excursion into entanglement measures
of concurrence-type we can state a connection between this type
of entanglement measures and lower bounds on the time
complexity for gate simulation.  The tilde mapping of
Def.~\ref{tilde}, which was used in 
the main body of the text, can be interpreted in the context
of concurrence-type entanglement measures. 
Since $H^{T}=H^{*}$ holds
for all Hermitian operators we obtain
\begin{equation*}
\Psi(U)=J_{n}' (\exp(\ii H))^{T} (J_{n}')^{-1}
=\exp(\ii J_{n}' H^{*} (J_{n}')^{-1} ),
\end{equation*}
where $\ii J_{n}' H^{*} (J_{n}')^{-1}$ is 
up to a minus sign equal to the rhs of  Eq.~(\ref{iH}).
This highlights that if we consider  the lower bounds 
introduced in Sec.~\ref{sectlow}
we are essential in setting of
Ref.~\cite{Uhl:2000} with $\vartheta=\vartheta_{1}$.

In both constructions, for the lower bounds on the time complexity
to  simulate unitary operators  and for the computation of the concurrence,
the essential point is that the spectrum
of both $U\tilde{U}$ and $\rho \tilde{\rho}$ is invariant under
local unitary operations. 
This highlights that there is a connection between
entanglement measures and lower bounds on the time
complexity for gate simulation. We are looking forward to generalizing
some of these ideas.

\section{Conclusion\label{conclusion}}
Starting with an extensive reconsideration of
infinitesimal Hamiltonian simulation and gate
simulation in the two-qubit case, we 
streamlined the different approaches 
by using Lie-theoretic methods.
As the success of this approach suggests,
this seems to be the appropriate level
of description for such a theory.

Going beyond two-qubits, we derived lower bounds
on the time complexity for gate simulation.
For this aim, we developed an  analogon
of the magic basis for general multipartite qubit-systems.
This gives us a first idea of the structure of unitary
operations w.r.t.\ the set of local unitary operations.
In addition, we related our approach to entanglement measures
of concurrence-type.
 
\begin{acknowledgments}
We are especially grateful to Dominik Janzing for many 
valuable discussions and comments.
In addition, we thank Steffen Glaser and  Thomas Schulte--Herbr\"uggen
for enlightening discussions.
The authors acknowledge the support of the Deutsche Forschungsgemeinschaft
(SPP ``Quanteninformationsverarbeitung'') under
Grant No. Be 887/13.
\end{acknowledgments}

\appendix*
\section{Spectral theory for infinitesimal Hamiltonian simulation\label{majHam}}
In this appendix  we prove a similar version of 
Thm.~\ref{factHsimmaj}.
Relying on Subsection~\ref{subsectmaj},
we use arguments from the theory of majorization applied 
to the spectrum of Hamilton operators.
We denote the vector of eigenvalues of the $k\times k$-dimensional matrix 
$A$, including multiplicities, by $\spec(A)
=(\spec(A)_{1},\ldots,\spec(A)_{k})^{T}$. In addition, we assume
that $\spec(A)_{i}\geq \spec(A)_{j}$ if $i<j$
($1\leq i,j \leq k$). The
majorization of the spectra of two matrices is, by a theorem of
Uhlmann~\cite{Uhl:1971}, related to
the convex combination of unitary orbits.
\begin{fact}[Uhlmann, see, e.g., {\cite[Satz~3]{Uhl:1971}} or
{\cite[Thm.~2-2.]{AU:1982}}\label{Uhlthm}] 
For Hermitian matrices $A$ and $B$ the condition 
$\spec(A) \prec \spec(B)$ 
is equivalent to
\begin{equation*}
A=\sum_{i} q_{i} U_{i}^{-1} B U_{i}, 
\end{equation*}
where $U_{i}$ are unitary
matrices,  $q_{i}\geq 0$, and  $\sum_{i} q_{i}=1$.
\end{fact}
We need another theorem connecting the notion of majorization
with the convex hull of all permuted versions of a vector.
\begin{fact}[Rado, see, e.g., 
Ref.~{\cite{Rad:1952}\label{Radthm}} 
or Ref.~{\cite[Prop.~4.C.1.]{MO:1979}}]
The vector $x$ is majorized by the vector $y$
if and only if
$x$ lies in the convex hull of all permutations
of $y$. 
\end{fact}

The spectral version of Thm.~\ref{factHsimmaj} reads:

\begin{thm}[{\cite[pp.~9--10]{VC:2002}}]
Assume that $H$ and $H'$ are non-local Hamilton operators
acting on a two-qubit system.
Let $a$ and $a'$ be  elements of $\ga$,
where
$a=a_{1}X_{7}+a_{2}X_{8}+a_{3}X_{9}=(\Ad_{\g}(L))(H)$
and 
$a'=a'_{1}X_{7}+a'_{2}X_{8}+a'_{3}X_{9}=(\Ad_{\g}(L'))(H')$
for some local unitary operators $L$ and $L'$.

A two-qubit system with Hamilton operator $H$ and local unitary
operators available is able to simulate the Hamilton operator $H'$ 
in time $t$ if and only if 
\begin{equation*}
\spec(a') \prec t\, \spec(a).
\end{equation*}
\end{thm}
\begin{rem}
The necessity of this condition was proven in 
Ref.~\cite{WJB:02}. In the prove we follow
Ref.~\cite{VC:2002}.
\end{rem}
\begin{proof}
Due to Fact~\ref{p char} we can choose $a$ and $a'$ as given.
The ``only if''-case follows by  Fact~\ref{Uhlthm}. We consider now
the ``if''-case. By invoking Fact~\ref{Radthm}, we get that 
\begin{equation*}
\spec(a'/t)=\sum_{k} q_{k} P_{k} \spec(a),
\end{equation*}
where $P_{k}$ is a permutation, $q_{k} \geq 0$, and 
$\sum_{k} q_{k} = 1$.
Since $a$ and $a'$ are elements of $\ga$, they commute. It follows that
there exists a basis where $a$ and $a'$ are simultaneously diagonal.
In that basis the permutations $P_{k}$ correspond
to permutations of the diagonal elements of $a$. For that reason we have that
\begin{equation*}
(a'/t) = \sum_{k} q_{k} U^{-1}_{k} a U_{k}
\end{equation*}
for some unitary operators $U_{k}$ which permute the
spectrum of $a$. We emphasize that the $U_{k}$
are not necessarily local. But we prove now that we can
find local unitary operators implementing any permutation
of the spectrum of $a$. Conjugation by the 
local unitary operators 
$
((\sigma_{0}-\ii \sigma_{1})/{\sqrt{2}})\otimes
((\sigma_{0}-\ii \sigma_{1})/{\sqrt{2}})
$,
$
((\sigma_{0}+\ii \sigma_{3})/{\sqrt{2}}) \otimes
((\sigma_{0}+\ii \sigma_{3})/{\sqrt{2}})
$,
and
$
((\sigma_{0}+\ii \sigma_{1})/{\sqrt{2}}) \otimes
((\sigma_{0}-\ii \sigma_{1})/{\sqrt{2}})
$
permutes the eigenvalues as respectively
follows:
$
(1,2,3,4) \mapsto (2,1,3,4)
$, 
$(1,2,3,4) \mapsto (1,3,2,4)
$, 
and
$(1,2,3,4) \mapsto (1,2,4,3)$.
As all permutations on four-vectors are generated by this permutations,
the ``if''-case follows.
\end{proof}
We note that the local unitary operators 
that permute the spectrum of elements of $\ga$
are given in
Ref.~\cite{VC:2002}, but there the second local unitary
operator is misprinted.


\end{document}